\documentclass[linenumbers]{aastex631}

\usepackage{setspace}

\newcommand{\bulge}{$F(G,M_{20})$}
\newcommand{\sersic}{Sérsic}
\newcommand{\age}{$\mathrm{age}_L$}
\newcommand{\grad}{$\nabla_{1R_e}$}
\newcommand{\cumQM}{$\Sigma E_{\mathrm{QM}}$}
\newcommand{\lomass}{$M_*<10^{10} M_{\odot}$}
\newcommand{\himass}{$M_*>10^{10.5} M_{\odot}$}
\newcommand{\sfgas}{$\Delta M_{\mathrm{SFG}}$}
\newcommand{\SFR}{$\rm \Delta \Sigma_{SFR}$}

\submitjournal{ApJ}

\begin{document}

\title{Intrinsic and Environmental Effects on the Distribution of Star 
Formation in TNG100 Galaxies}

\correspondingauthor{Bryanne McDonough}
\email{br.mcdonough@northeastern.edu}

\author[0000-0001-6928-4345]{Bryanne McDonough}
\affiliation{Department of Physics, Northeastern University, 360 Huntington Ave, Boston, MA}
\affiliation{Institute for Astrophysical Research, Boston University, 725 Commonwealth Ave, Boston, MA}%https://www.overleaf.com/project/6397926f74163487fc461872

\author[0000-0002-0212-4563]{Olivia Curtis}
\affiliation{Department of Astronomy \& Astrophysics, The Pennsylvania State University, 251 Pollock Road, University Park, PA}
\affiliation{Institute for Astrophysical Research, Boston University, 725 Commonwealth Ave, Boston, MA}

\author[0000-0001-7917-7623]{Tereasa G. Brainerd}
\affiliation{Institute for Astrophysical Research, Boston University, 725 Commonwealth Ave, Boston, MA}

%% Note that the \and command from previous versions of AASTeX is now
%% depreciated in this version as it is no longer necessary. AASTeX 
%% automatically takes care of all commas and "and"s between authors names.

%% AASTeX 6.31 has the new \collaboration and \nocollaboration commands to
%% provide the collaboration status of a group of authors. These commands 
%% can be used either before or after the list of corresponding authors. The
%% argument for \collaboration is the collaboration identifier. Authors are
%% encouraged to surround collaboration identifiers with ()s. The 
%% \nocollaboration command takes no argument and exists to indicate that
%% the nearby authors are not part of surrounding collaborations.

%% Mark off the abstract in the ``abstract'' environment. 
\begin{abstract}
We present radial profiles of luminosity-weighted age (\age) \textnormal{and \SFR{} }for various populations of high- and low- mass central and satellite galaxies in the TNG100 cosmological simulation. Using these profiles, we investigate the impact of intrinsic and environmental factors on the radial distribution of star formation. For both central galaxies and satellites, we investigate the effects of black hole mass, cumulative AGN feedback energy, morphology, halo mass, and local galaxy overdensity on the profiles. In addition, we investigate the dependence of \textnormal{radial} profiles of the satellite galaxies as a function of the redshifts at which they joined their hosts, as well as the net change in star-forming gas mass since the satellites joined their host. We find that high-mass (\himass) central and satellite galaxies show evidence of inside-out quenching driven by AGN feedback. Effects from environmental processes only become apparent in averaged profiles at extreme halo masses and local overdensities. We find that the dominant quenching process for low-mass galaxies (\lomass) is environmental, generally occurring at low halo mass and high local galaxy overdensity for low-mass central galaxies and at high host halo masses for low-mass satellite galaxies. Overall, we find that environmental processes generally drive quenching from the outside-in. 
%The net effect of environmental processes is a positive slope in \age{} at large radii ($R \gtrsim 1 R_e$). Quenched, low-mass (\lomass) central galaxies are rare, but tend to exist at high local galaxy overdensities and low halo masses. In contrast, most quenched low-mass satellite galaxies are found in high mass halos.

\end{abstract}

%% Keywords should appear after the \end{abstract} command. 
%% The AAS Journals now uses Unified Astronomy Thesaurus concepts:
%% https://astrothesaurus.org
%% You will be asked to selected these concepts during the submission process
%% but this old "keyword" functionality is maintained in case authors want
%% to include these concepts in their preprints.

%% From the front matter, we move on to the body of the paper.
%% Sections are demarcated by \section and \subsection, respectively.
%% Observe the use of the LaTeX \label
%% command after the \subsection to give a symbolic KEY to the
%% subsection for cross-referencing in a \ref command.
%% You can use LaTeX's \ref and \label commands to keep track of
%% cross-references to sections, equations, tables, and figures.
%% That way, if you change the order of any elements, LaTeX will
%% automatically renumber them.
%%
%% We recommend that authors also use the natbib \citep
%% and \citet commands to identify citations.  The citations are
%% tied to the reference list via symbolic KEYs. The KEY corresponds
%% to the KEY in the \bibitem in the reference list below. 

\section{Introduction} \label{sec:intro}

The advent of integral field spectroscopy (IFS) surveys has revolutionized the study of star formation on resolved scales outside of the Milky Way, \textnormal{leading to new insights into how galaxies grow and evolve. Here, we briefly describe these insights as they relate to the radial distribution of properties related to star formation; we direct readers to \cite{Sanchez20} and \cite{Sanchez21} for a comprehensive review. At the onset of star formation, the gas reservoir of galaxies is centrally concentrated. Over time, stars form out of molecular gas at larger and larger radii. This results in the negative gradient in the stellar mass surface density that is seen at present-day \citep[e.g.,][]{Gonzalez14,Sanchez20,Pan24}. This gradient, sometimes referred to as ``inside-out growth'' is more extreme in more massive/early-type galaxies.} 

\textnormal{The cessation of star formation also appears to occur from the inside-out, at least for massive galaxies.
This is reflected by negative gradients in the luminosity-weighted age of massive galaxies over large scales ($R/R_e \approx 0- 3$), e.g., \cite{Gonzalez14,Zheng17,Sanchez20}.}
The central suppression of \textnormal{specific star formation rates (sSFR), or similar parameters}, in massive galaxies ($M_*/M_{\odot} \gtrsim 10^{10}$) in the observed universe has been well documented in the literature \citep[e.g.,][]{Gonzalez14,Ellison18,Spindler18,Bluck2020b,Sanchez20,Sanchez21,Pan24}.
%This inside-out quenching is thought to be driven by the central supermassive black hole (SMBH), due to 
The tight correlation between the mass of the central supermassive black hole (SMBH) and other galaxy properties, \textnormal{including stellar mass and star formation rate (SFR), indicates that SMBHs play a significant role in quenching massive galaxies}. \textnormal{In the parlance of \cite{Peng10}, `mass quenching' refers to processes inherently linked to the mass of a galaxy that act to quench star formation, leading to larger fractions of quenched galaxies with increasing stellar mass and SMBH mass \citep[e.g.,][]{Peng10, Bluck2020a}}. \cite{Bluck2020b} used a random forest machine learning algorithm to determine the importance of various parameters for predicting quenching in central galaxies observed by MaNGA \citep{MaNGA}. They found that central velocity dispersion, which is tightly correlated with SMBH mass \citep[e.g.,][]{Saglia}, was the most predictive variable. A SMBH can regulate star formation via AGN energy imparted onto cool gas.
%, resulting in heating or removal.
\textnormal{The exact mechanisms by which AGN feedback results in a cessation of star formation is still an active area of research. However, it may be the result of heating or removal of gas reservoirs, or the introduction of turbulence, resulting in a decrease in the star formation efficiency (SFE). SFE has been found to decrease during the `green valley' stage of galaxy evolution, the transition from actively star-forming to retired/quenched \citep[e.g.,][]{Colombo20,Villanueva24}. The cause of decreasing SFE and its role in shutting down star formation is still being explored.}

Determining which parameters drive quenching in galaxies is complicated by the inter-correlation of variables, particularly those that are intrinsic to a galaxy. This may be especially true for parameters that characterize galaxy morphology. For example, \cite{Bluck2020b} found that bulge-to-total stellar mass ratio (B/T) was the fourth most important parameter for quenching in central galaxies, but they note that its importance may be overestimated due to the correlation of B/T with central velocity dispersion. In addition, \cite{Bluck14} identified bulge mass as being tightly correlated with the quenched fraction of galaxies.

The role of morphology in shaping the radial distribution of stellar ages and local sSFR has also been investigated by \cite{Gonzalez14,GonzalezDelgado16} in the CALIFA survey \citep{CALIFA}. \cite{Gonzalez14} found that the shape of the average luminosity-weighted age (\age) profile varied as a function of stellar mass and concentration index.
%, and concluded that the \age{} profiles of spheroidal galaxies were flatter than the profiles of disk-dominated galaxies. \cite{Gonzalez14} found that the largest negative age gradient in the inner regions of the \age{} profiles occurred for massive disk galaxies. Additionally, \cite{GonzalezDelgado16} examined radial profiles of sSFR as a function of morphological type. From this, \cite{GonzalezDelgado16} found that galaxies that are dominated by a spheroidal component were completely quenched, while disk-dominated galaxies were actively forming stars, resulting in local sSFRs that increased with increasing distance from the centers of the galaxies. 
\cite{Martig} proposed that the presence of a bulge can stabilize gas against the gravitational collapse and fragmentation necessary for star formation. It is also possible that the presence of a bulge and quiescence can be the result of similar evolutionary processes; e.g., mergers. Due to the correlation of morphology with stellar mass, the role of morphology in galaxy quenching remains unclear.

The halo environment in which a galaxy resides can also affect star formation properties via processes that remove, heat, or compress gas, such as galaxy-galaxy harassment, ram pressure stripping, and dynamical stripping \citep[see, e.g.,][]{Peng12}. \textnormal{\cite{Corcho-Cabellero2023} found that most galaxies that quenched relatively rapidly (over $\sim 1$ Gyr timescales) were low-mass satellite galaxies.}
%a sentence citing various studies showing effect of environment on star formation, including some cited in Bluck but also Ghodsi and Vulcani https://academic.oup.com/mnras/advance-article/doi/10.1093/mnras/stae279/7590826 CITE
The random forest analysis performed by \cite{Bluck2020b} indicated that, unlike central galaxies, both intrinsic and environmental parameters are predictive of quenching in satellite galaxies. In their work, \cite{Bluck2020b} found that environmental parameters, namely local galaxy overdensity and host halo mass, were most important for predicting quenching in low-mass ($M_*/M_{\odot}<10^{10}$) satellite galaxies. This boundary, $M_*/M_{\odot}\approx10^{10}$, has been indicated as the threshold above which quenching can be driven by a central AGN, also known as mass quenching \citep{Peng10,Peng12}. \cite{McDonough23} found a similar threshold at $10^{10}<M_*/M_{\odot}<10^{10.5}$ for galaxies in the TNG100 simulation, with more massive galaxies showing clear evidence of inside-out mass quenching in radial profiles of \age. 

Complimentary to the progress \textnormal{in spatially resolving galaxies in observations}, cosmological hydrodynamic simulations have improved in both scale and resolution, as well as the degree of agreement between observations and the predictions of the simulations. Key to the improvement in agreement between observations and simulations has been the implementation and improvement of feedback models in the simulations. These feedback models approximate the impact of active galactic nucelei (AGN) and supernovae, decreasing the efficiency of star formation and preventing early depletion of gas reservoirs \citep[see, e.g.,][]{Naabreview, Donnari19}. However, since the actual processes that drive AGN and supernovae feedback occur on scales that are not resolved in most simulations, these effects must be approximated, requiring assumptions regarding how and where feedback energy is imparted in simulated galaxies. 

In \cite{McDonough23}, we demonstrated that TNG100, part of the IllustrisTNG cosmological magnetohydrodynamic simulation suite \citep{Nelson2018,2018MNRAS.475..648P,2018MNRAS.475..676S,2018MNRAS.477.1206N,2018MNRAS.480.5113M,Nelson2019a}, reproduced both the observed resolved star formation main sequence (rSFMS) and the radial distribution of star formation in main sequence, green valley, and quenched galaxies. \textnormal{The rSFMS, or $\Sigma_* -\Sigma_{\mathrm{SFR}}$ relation, was first presented by \cite{Sanchez13} and \cite{Wuyts13} and was later characterized by \cite{Cano-Diaz16}.}
As demonstrated by \cite{Nelson21}, the location within galaxies at which feedback energy is injected is critical for accurately reproducing the observed radial distribution of star formation. Both TNG100 and its predecessor, Illustris-1, adopted AGN feedback models with prescriptions for high- and low- accretion rate modes. Illustris-1 \citep{Vogelsberger14,Nelson15} adopted an AGN feedback model that, at low accretion rates, injected feedback energy as thermal energy in large bubbles displaced from central galaxies. In contrast, the low-accretion state of the AGN feedback model in the IllustrisTNG simulations \citep{Weinberger17} injects kinetic energy into gas immediately surrounding the supermassive black hole at the galaxy center.

\cite{Nelson21} computed the radial profiles of sSFR for galaxies in the $50$~Mpc simulation box of IllustrisTNG (TNG50), Illustris-1, and 3D-HST observations.  From this, they found that the observed central suppression of star formation in massive galaxies was reproduced in TNG50 galaxies, but not in Illustris-1 galaxies. \cite{Nelson21} attribute this improvement to the difference in feedback models, since the TNG and Illustris-1 simulations are similar in other respects, including the same initial conditions and star formation prescriptions.

\textnormal{In this article, we adopt a broad but commonly used definition of `quenching,' which we use to describe the reduction and eventual cessation of star formation in galaxies. In this parlance, `quenched' galaxies are those that have ceased forming stars at significant levels, often defined by their position on a color-magnitude or $M_*- {\rm SFR}$ diagrams.
%Quenched galaxies may retain some degree of star formation, although this can be difficult to diagnose in both observations and simulations due to limited resolutions. 
We note that there is some discussion in the literature \citep[e.g.,][]{Corcho-Cabellero2023} that `quenching' be reserved to describe rapid cessation of star formation. \cite{Corcho-Cabellero2023} argues for a distinction between galaxies that quench rapidly and those that `retire' their star formation over longer timescales. While we adopt a broad definition of quenching in this article, we acknowledge that such a distinction may be relevant to future analyses. }

In this paper, we investigate the dependence of radial profiles of luminosity-weighted age and \textnormal{\SFR{}, the offset from the resolved star formation main sequence (rSFMS, or the $\Sigma_* - \Sigma_{\rm SFR}$ relation),} on various intrinsic and environmental parameters. The paper is organized as follows. In \S~\ref{sec:methods}, we provide a brief overview of TNG100, along with a discussion of the ways in which we derive parameters from the simulation data. In \S~\ref{sec:inprof}, we present results for radial profiles of \age{} \textnormal{and \SFR{} }for galaxies as a function of various intrinsic parameters. In \S~\ref{sec:envprof}, we explore the dependence of radial profiles of \age{} \textnormal{and \SFR{}} on environmental parameters. A discussion of the implications of our results, in the context of the existing literature, is presented in  \S~\ref{sec:discussion}, and a summary of our main results and conclusions is given in \S~\ref{sec:summary}. In Appendix \ref{app_corr}, we present correlations that exist between the various parameters that we explored in the main body of the paper. \textnormal{In Appendix \ref{sec:app_summary}, we summarize the normalization and gradients for the radial profiles presented in \S~\ref{sec:inprof} and \S~\ref{sec:envprof}. In Appendix~\ref{sec:app_mass}, we explore the mass-dependence of the profile normalizations.}

\section{Methodology} \label{sec:methods}

The methodology by which we select our galaxy sample is identical to the methodology we used in \cite{McDonough23}.  That is, 
a sample of $\sim 60,000$ luminous ($M_r < -14.5$) simulated galaxies at $z = 0$ was obtained from the $75^3 h^{-3}\rm{Mpc}^3$ volume TNG100-1 simulation (hereafter, TNG100). The TNG100 simulation is part of the IllustrisTNG project, a suite of publicly available $\Lambda$CDM magnetohydrodynamical simulations that adopted a \citet{Planck2016} cosmology with the following parameter values:
$\Omega_{\Lambda,0}=0.6911$, $\Omega_{m,0}=0.3089$, $\Omega_{b,0}=0.0486$, $\sigma_8=0.8159$, $n_s=0.9667$, and $h=0.6774$.  We take the galaxy stellar mass to be the total mass of all stellar particles that are bound to each galaxy according to the \texttt{SubFind} catalog. From the Stellar Projected Sizes supplementary catalog \citep{Genel2018}, we obtain $r$-band half-light radii, which we will refer to as effective radii, $R_e$. We obtain magnitudes from the SDSS Photometry, Colors, and Mock Fiber Spectra catalog \citep{Nelson2018}. It is also important to note that, unlike $R_e$ and galaxy magnitudes (which are directly comparable to observed galaxy properties), the stellar masses we adopt for TNG100 galaxies are not directly comparable to the stellar masses of observed galaxies (which are typically obtained by modelling of the spectral energy distribution).

Here, galaxy sample and data products are identical to those that we used in \cite{McDonough23}.  We provide only a brief description here, and the reader is referred to Section 2 of \cite{McDonough23} for a full description of the catalogs and galaxy sample selection criteria. We ensure sufficient resolution of the galaxies by limiting the sample to galaxies that contain at least $1,000$ stellar particles within $2R_e$ and have a physical size of $R_e > 4$~kpc. The most massive galaxy in a given friends-of-friends group is defined to be the central galaxy, and all other group members are considered to be satellites. Our full sample consists of $4,605$ central galaxies and $1,592$ satellites. From \cite{McDonough23}, we know that the shape\textnormal{s} of the \textnormal{radial} profile\textnormal{s} differ significantly for high-mass and low-mass galaxies, and we define these to be \himass{} and \lomass, respectively. The division of our sample into low- and high-mass galaxies is based on Figure 5 of \cite{McDonough23} and, under this definition, our sample contains $2,167$ high-mass galaxies and $2,709$ low-mass galaxies. 

In Sections \ref{sec:inprof} and \ref{sec:envprof}, we explore the role of various parameters on the radial distribution of \age{} \textnormal{and \SFR{}} within galaxies. We do this by constructing radial profiles  in the same manner as \cite{McDonough23}, and the reader is referred to Section 3.4 of \cite{McDonough23} for full details of our methods. In brief, \textnormal{\age{} profiles are constructed by directly projecting the stellar particles of each galaxy into a 2D, face-on orientation, and radial bins are normalized by the effective radius of the galaxy.} A luminosity-weighted age is computed using all particles in a given radial bin for a given galaxy population. \textnormal{In the case of \SFR{} profiles, we construct face-on maps of stellar-mass density, $\Sigma_*$, and the density of SFR, $\Sigma_{\rm SFR}$, by smoothing stellar particles onto a 2D grid using a cubic spline kernel. Maps of $\Sigma_{\rm SFR}$ are constructed using a combination of stellar particles formed in the last $20$ and $100$~Myr, timescales that are roughly comparable to the H$\alpha$ and D$4000$ observational tracers of star formation. Additionally, a fixed sSFR of $10^{-12} {\rm yr}^{-1}$ is assumed for spatial bins where star formation was not resolved in the simulation. Finally, a map of \SFR{} is constructed by determining the logarithmic offset of a given spatial bin from the resolved star-forming main sequence presented in \cite{McDonough23}. Our \SFR{} profiles were constructed to be comparable to the \SFR{} profiles of MaNGA galaxies presented in \cite{Bluck2020b}. }

In addition, we compute gradients, $\nabla_{1R_e}$, for the profiles (which are taken to be the slope of a line fit to the profiles in the radius range $0 \le r \le 1 R_e$), and we only present profiles for galaxy populations when there are \textnormal{at least $10$} representative objects in our sample. \textnormal{We note that the aging of stellar populations that formed during the inside-out growth of galaxies can naturally result in negative gradients in \age{}. However, the luminosity-weighting will bias the measured \age{} toward more recent star formation.} \textnormal{While other works \citep[e.g.,][]{GonzalezDelgado16,Sanchez21} measure gradients over different regimes and measure the profiles out to larger radii, we have here tailored our derivation of \SFR{} for comparison to \cite{Bluck2020b}, who used a single gradient and were limited in radial range by the coverage of the primary MaNGA sample.}

\textnormal{While \SFR{} traces star formation activity over the last $100$ Myr, \age{} reflects both recent star formation activity (emphasized via the luminosity-weighting) and the natural aging of stellar populations. Thus, \age{} profiles will be useful for understanding quenching processes that operate over longer timescales, while \SFR{} will be more useful for identifying parameters that contribute to rapid quenching. As reported in \cite{McDonough23}, profiles of \SFR{} in galaxies undergoing active quenching are biased toward low \SFR{} at intermediate radii, likely due to resolution limits imposed by the simulation. }

In Appendix \ref{app_corr}, we explore the interdependence of the parameters that we discuss in this paper, and we do this separately for central and satellite galaxies. \textnormal{In Appendices \ref{sec:app_summary} and \ref{sec:app_mass}, the normalization of \age{} profiles are taken to be the total \age{} of a galaxy using all stellar particles within $1.5R_e$ and the normalization of \SFR{} profiles are taken to be the logarithmic offset of a galaxy's SFR from the star-forming main sequence identified for TNG100 galaxies in \cite{McDonough23}.}
%In Appendix \ref{sec:app_summary}, we summarize the profile gradients and global values of \age{} and the logarithmic offset from the global star-forming main sequence identified in \cite{McDonough23}, $\Delta$SFR. In Appendix C, we explore the mass-dependency of the global \age{}

\subsection{Intrinsic Parameters} \label{sec:methodIP}
In Section \ref{sec:inprof}, we present the ways in which intrinsic factors affect the radial distribution of luminosity-weighted stellar age. There we show population-averaged profiles for our sample, subdivided by supermassive black (SMBH) mass, cumulative feedback energy injected by the SMBH in the high-accretion rate (`quasar') mode, as well as the morphological Gini-$M_{20}$ `bulge statistic' \citep{Snyder2015}, which separates early- and late- type galaxies by quantifying the relative dominance of galaxy bulges. In addition to the results from these three parameters, we also briefly summarize results we obtain for other parameters that are related to the SMBH and morphology.

From the TNG100 \texttt{SubFind} catalog, which identifies parameters for individual subhalos (i.e., galaxies), we obtained the $z=0$ SMBH masses, $M_{BH}$, and instantaneous SMBH accretion rates for our sample. 
Since the instantaneous accretion rate is not indicative of processes that affected the $z = 0$ radial profiles, we also obtained the SMBH accretion rate from the penultimate simulation snapshot using the \texttt{Sublink} merger trees, which corresponds to the SMBH accretion rate at redshift $z \approx 0.01$. We note that, in TNG100,
black hole particles are seeded into galaxies when the halo mass exceeds $5\times 10^{10} h^{-1}  M_\odot$, and the galaxy does not already contain a SMBH. This black hole seeding prescription results in the majority of galaxies in our sample having SMBHs. However, a small number ($107$), most of which are low-mass satellite galaxies, lack SMBHs. 

We obtained the cumulative feedback energy imparted by a black hole on its host galaxy, $\Sigma E_{\mathrm{AGN}}$, directly from the particle data. To do this, we identified the most massive black hole particle present within the effective radius of each galaxy in the $z=0$ snapshot. These particles lie at the centers of the galaxies, since the TNG algorithm employs a centering prescription that ensures the SMBHs remain at the gravitational potential minima of their hosts' halos. Each black hole particle contains several tracked fields, including cumulative feedback energy injected in both the quasar mode and radio mode. The quasar (or 'thermal') mode operates when the black hole particle is in the high accretion state, and thermal energy is injected into the surrounding gas. The radio (or 'kinetic') mode operates when the particle is in the low accretion state, in which case pure kinetic energy is injected into the surrounding gas. In our sample, $3,598$ galaxies experienced energy being injected via the quasar mode ($\Sigma E_{\mathrm{QM}}$) during their lifetimes. Only $337$ galaxies in our sample experienced injection of kinetic energy ($\Sigma E_{\mathrm{RM}}$) during their lifetimes. Such a low number of galaxies with non-zero values of $\Sigma E_{\mathrm{RM}}$ is not necessarily surprising since the radio mode generally takes over at low redshift in galaxies with $M_* > 10^{10.5} M_\odot$ \citep{Weinberger18}. In Section \ref{sec:SMBH}, we show profiles of galaxy populations subdivided by cumulative energy imparted in the quasar mode, as that is the dominant, more energetic process. 
However, we also briefly discuss profiles of galaxy populations that are subdivided by cumulative energy imparted in the radio mode and the total overall energy imparted, $\Sigma E_{\mathrm{AGN}}$, where $\Sigma E_{\mathrm{AGN}} = \Sigma E_{\mathrm{QM}} + \Sigma E_{\mathrm{RM}}$.

Below, we also explore the dependence of radial profiles on morphology using the Gini-$M_{20}$ bulge statistic. The SKIRT Synthetic Images and Optical Morphologies supplementary catalog from \cite{Rodriguez-Gomez19} contains morphological parameters for $4999$ galaxies in our sample at $z=0$, including the bulge statistic. We also briefly discuss results using the Sérsic index \citep{Sersic63}, $n$, which was obtained from 2D Sérsic fits to synthetic $i$-band images and is provided in the \cite{Rodriguez-Gomez19} supplementary catalog.

The Gini-$M_{20}$ bulge statistic, hereafter \bulge, is defined in \cite{Snyder2015} \textnormal{and describes the location of a galaxy on a diagram plotting the Gini coefficient ($G$) against the M$_{20}$ statistic. The Gini coefficient, traditionally used as a statistic in economics to study the distribution of wealth, was proposed as a method to measure the distribution of flux in pixels by \cite{Abraham03} and \cite{Lotz04}. At $G=1$, all of the flux from a galaxy is concentrated in a single pixel, while at $G=0$, the flux is spread evenly among every pixel. M$_{20}$ measures the second-order moment of the brightest $20\%$ of the flux in a galaxy.} 

\textnormal{The combination of $G$ and M$_{20}$ into the \bulge{}} statistic separates galaxies into early- and late- types, and indicates the relative dominance of the bulge. The greater the dominance of the bulge, the higher the value of \bulge{} is above zero. The more disk-dominated a galaxy is, the lower the value of \bulge{} is below zero. The bulge statistic is strongly correlated with other statistics that measure galactic bulge strength, including the \sersic{} index and concentration. However, \bulge{} is less sensitive to mergers and other disturbances, making it a better indicator of bulge strength in our satellite sample.

\subsection{Environmental Parameters}

For both central and satellite galaxies, we explore the dependence of radial \age{} profiles on halo mass and local galaxy overdensity. 
%Details of the calculation of \age{} profiles can be found in \cite{McDonough23}. - said above
We take the halo mass to be the total mass of all particles belonging to the friends-of-friends group to which a galaxy belongs. To best compare our results for simulated galaxies to the observational results obtained \cite{Bluck2020b} for observed galaxies, we evaluate the local galaxy overdensity at the fifth nearest neighbor with an absolute $r$-band magnitude of $M_r<-20$. The local galaxy overdensity is then computed according to:

\begin{equation}
    \delta_5 \equiv \frac{n_5}{\bar{n}}-1 = \frac{5}{\frac{4}{3} \pi d_5^3} \frac{L_{\mathrm{box}}^3}{N_{\mathrm{total}}}-1,
\end{equation}
where $d_5$ is the distance to the fifth nearest neighbor, $N_{\mathrm{total}}$ is the total number of TNG100 galaxies with $M_r<-20$, and $L_{\mathrm{box}}$ is the box length of the simulation. 

\subsubsection{Satellite Parameters}

In satellite galaxies, interactions \textnormal{with} other galaxies \textnormal{and with the hot gas in their hosts' halos} can affect the availability of gas for star formation. It is not trivial to parameterize these interactions; e.g., the degree to which a satellite has been affected by ram pressure stripping or galaxy-galaxy harassment. However, we can quantify how properties of each satellite have changed since it joined its host's halo. 

We define the satellite joining time to be the redshift at which the satellite first entered within $3R_{200}(z)$ of the center of its $z=0$ host halo. Here, $R_{200}(z)$ is the virial radius of the host galaxy at a given redshift. To obtain the positions and properties of galaxies and their $z=0$ groups, we used the \texttt{SubLink} merger trees of the galaxies. For the halo properties, we used the merger tree of the primary (most massive) group member at $z=0$. 

From the merger trees, we obtain the redshift at which the satellite joined ($z_j$), the redshift of the snapshot at which the satellite most recently approached perigalacticon ($z_{\mathrm{p}}$), and the separation between the satellite and its host at perigalacticon ($d_{h-s}$). 
%change in the kinetic energy of the satellite relative to the group center of mass velocity since joining, $\Delta T$. Satellite galaxies will lose kinetic energy as they experience dynamical friction in their orbit, so $\Delta T$ acts as a proxy to measure the effects of dynamical friction on star formation in satellites. 
Additionally, we identify the total change in the mass of bound, star forming gas since the satellite joined ($\Delta M_{\mathrm{SFG}}$). The gas particle data was obtained from the satellite subhalo cutouts, from which we computed the total mass of gas particles with non-zero instantaneous star formation rates. The difference in mass from the $z=0$ snapshot cutouts and the cutout at the snapshot corresponding to $z_j$ then yields the values $\Delta M_{\mathrm{SFG}}$ for a given galaxy.

In Section \ref{sec:envprof}, we present radial profiles of \age{} \textnormal{and \SFR{}} for galaxy populations subdivided by $z_j$ and \sfgas{}, and we briefly discuss results for $z_{\mathrm{p}}$ and $d_{h-s}$. 

\section{Dependence on Intrinsic Parameters} \label{sec:inprof}
\subsection{Supermassive Black Hole} \label{sec:SMBH}

%\subsubsection{Mass and Accretion Rate}
\begin{figure}
    \centering
    \includegraphics[width=\linewidth]{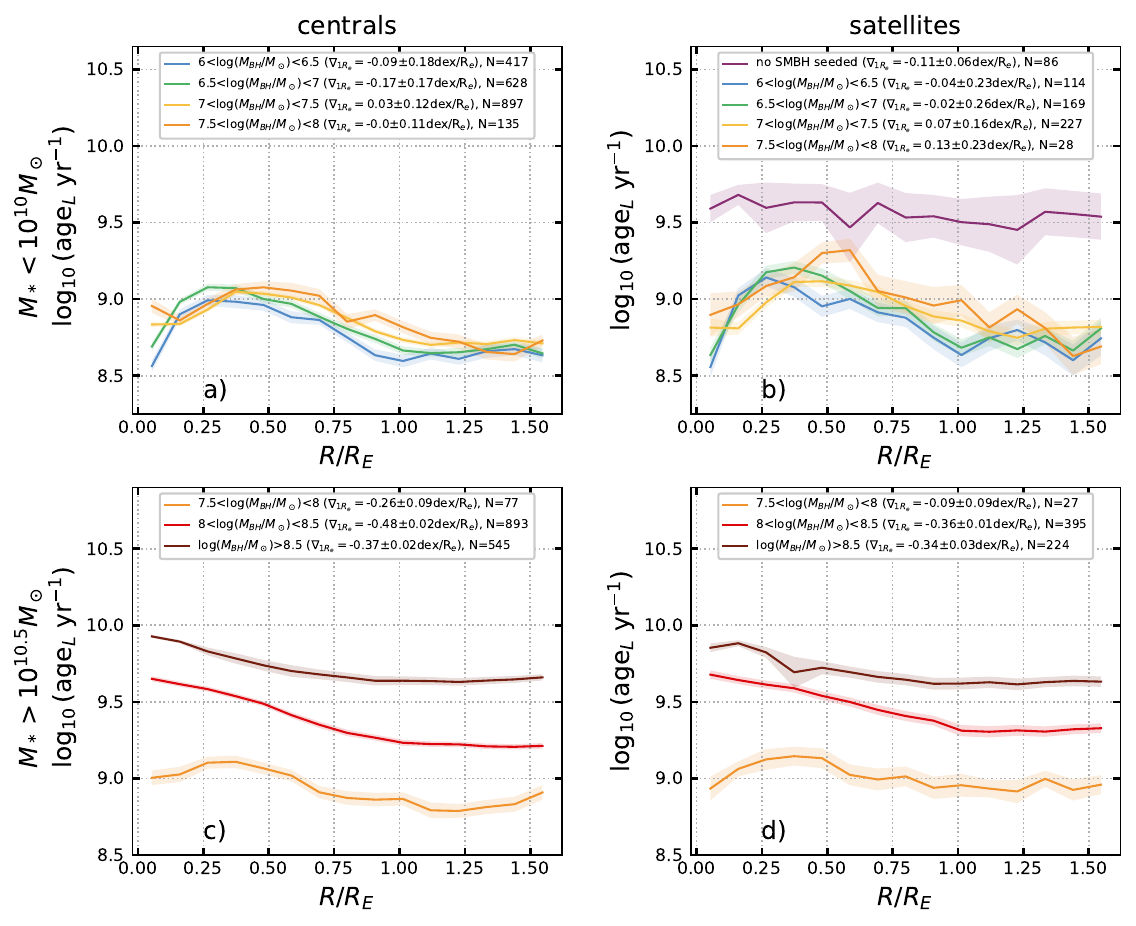}
    \caption{Radial profiles of \age{}, subdivided by SMBH mass. The legends indicate the profile gradients (\grad) and the number of galaxies that contributed to the profile ($N$). Profiles were computed separately for populations of high-mass (top row) and low-mass (bottom row) central (left column) and satellite  (right column) galaxies. Errors were computed using 2,000 bootstrap resamplings of the data. Error bounds that are wider than the line width are indicated by shaded regions and are omitted from the figure when they are comparable to or smaller than the line width.}
    \label{fig:BHmass}
  
\end{figure}

\begin{figure}
    \centering
    \includegraphics[width=\textwidth, keepaspectratio]{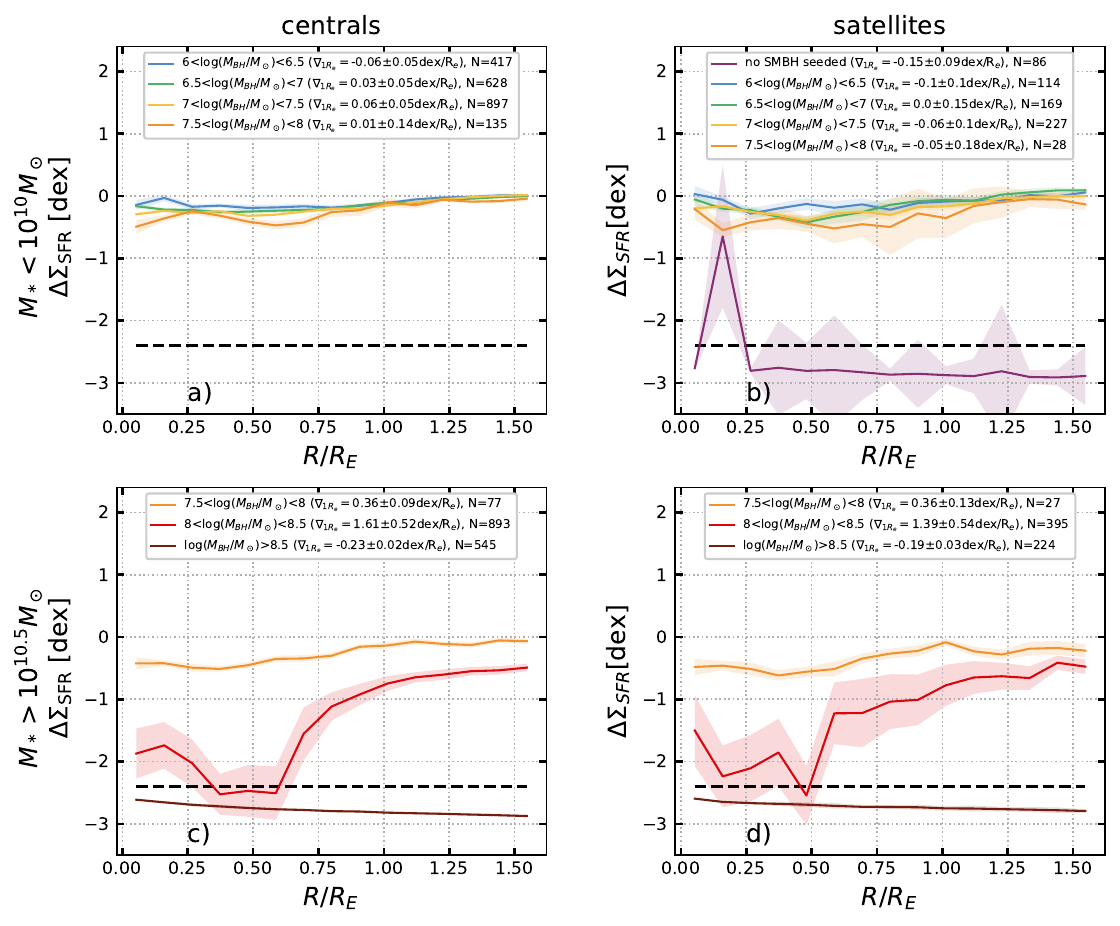}
    \caption{\textnormal{Radial profiles of \SFR{}, subdivided by SMBH mass. Here, the formatting is identical to that of Figure \ref{fig:BHmass}, but with the addition of a black dashed line that indicates the threshold below which a spatially resolved bin is fully quenched.}}
    \label{fig:SFR_BHMass}
  
\end{figure}

In Figures \ref{fig:BHmass} \textnormal{and \ref{fig:SFR_BHMass}}, we present profiles of luminosity-weighted age\textnormal{ and \SFR{}, respectively,} for populations of galaxies in our sample, subdivided by SMBH mass. The left panels show profiles for central galaxies, while the right panels show profiles for satellite galaxies. The top panels show profiles for galaxies with \lomass, while the bottom panels show profiles of high-mass galaxies (\himass). The boundaries for the SMBH mass bins, the profile gradients, and the number of galaxies in a given bin are included in the legends. Error bounds (computed via 2,000 bootstrap resamplings) are indicated by shaded regions when they are larger than the line widths, and are omitted when they are comparable to or smaller than the line widths. \textnormal{In Figure \ref{fig:SFR_BHMass}, the black dashed line indicates the threshold below which $\Sigma_{\rm SFR}$ is unresolved and is therefore artificially imposed.}

\textnormal{Purple} lines in Figures \ref{fig:BHmass} and \ref{fig:SFR_BHMass} give the profiles for galaxies that lack SMBHs (i.e., no SMBH was seeded in these galaxies; see Section \ref{sec:methodIP}). For low-mass satellites that lack SMBHs, the \age{} profiles are $\sim0.5$ dex ($\sim500$ Myr) older than the profiles of low-mass galaxies that do contain SMBHs. \textnormal{While there is a high degree of variability, \SFR{} profiles of low-mass satellites without SMBHs appear almost completely quenched, on average. There is some evidence that ongoing star formation occurs in the centers of these galaxies, although this is not reflected in the complimentary \age{} profiles. }
%For the $5$ high-mass satellite galaxies that lack SMBHs, the \age{} profile is close to that of galaxies with SMBH masses $M_{\mathrm{BH}}/M_\odot> 10^{8}$. 
The older \age{}\textnormal{ and quenched \SFR{} profiles} of satellites without SMBHs is consistent with previous results that find that the presence of a SMBH prevents overcooling and leads to overly-efficient early star formation \citep[e.g.,][]{Silk12}.

In the case of galaxies for which SMBHs were seeded, we find remarkable similarity in both shape and normalization of the \textnormal{\age{} and \SFR{}} profiles of low-mass galaxies with varying $M_{\mathrm{BH}}$. However, there is a clear difference in shape and normalization of profiles for high-mass galaxies with different values of $M_{\mathrm{BH}}$. That is, the more massive a SMBH, the older the \age{} profile \textnormal{and the lower the \SFR{} profile}. Feedback from SMBHs drives inside-out quenching \citep{Nelson21}, which is indicated by a negative \textnormal{(positive)} \age{} \textnormal{(\SFR)} profile gradient. Of the three seeded populations of high-mass central galaxies, the middle population, $10^{8}<M_{\mathrm{BH}}/M_\odot<10^{8.5}$, has the steepest slopes (\grad\age $= -0.48 \pm 0.02$ dex/$R_e$, \grad\SFR$=1.61 \pm 0.52$ dex/$R_e$). 

For both high-mass centrals and satellites, the $10^{7.5}<M_{\mathrm{BH}}/M_\odot<10^{8}$ \textnormal{(orange)} population has the \age{} profiles with the shallowest gradient. \textnormal{In the \SFR{} profiles (Figure \ref{fig:SFR_BHMass}), the profiles of high-mass galaxies with $M_{\rm BH}/M_\odot >10^{8.5}$ (purple) are completely quenched. By construction, such low values of \SFR{} are artificially imposed, so no further conclusions can be drawn from these profiles.}

While not shown here, we have also explored radial profiles of galaxies subdivided by the accretion rates of their SMBHs in both the $z=0$ snapshot and the previous snapshot, $z\approx 0.0095$. From this, we find that there is little difference between profiles based on accretion rates taken at different snapshots. Similar to $M_{\mathrm{BH}}$ above, the SMBH accretion rates have little effect on the shape or normalization of radial profiles of low-mass galaxies. \textnormal{There is a slight exception in the \SFR{} profiles of low-mass satellite galaxies with $\dot{M}_{\rm BH}^{z\approx 0.01}<10^{-5} M_\odot {\rm yr}^{-1}$. This population is highly variable, but on average, have lower values of \SFR{} than satellites with higher accretion rates.} The normalization of \age{} profiles of high-mass galaxies is correlated with accretion rate, but not to the same degree as profiles subdivided by $M_{\mathrm{BH}}$. These results are not surprising, as the feedback energy imparted by the accretion at $z=0$ will not have had time to affect the galaxy, and \age{} will be affected by feedback that is imparted over timescales longer than the time between $z=0$ and $z\lessapprox0.0095$. This is consistent with the results of \cite{Bluck23}, who found that quenching of star formation showed little dependence on instantaneous SMBH accretion rate, but strong dependence on total SMBH mass.

The amount of AGN feedback energy imparted onto a galaxy at a given snapshot is proportional to the instantaneous accretion rate of the black hole. Thus, the cumulative energy imparted will be proportional to the total accretion onto the black hole over its lifetime. Since the $z=0$ black hole mass is a summation of the matter accreted over its lifetime, $M_{\mathrm{BH}}$ and \cumQM{} will be well-correlated, and this is shown in Appendix \ref{app_corr}.

%\subsubsection{Cumulative Feedback Energy} \label{sec:cumE}
\begin{figure}
    \centering
    \includegraphics[width=\linewidth]{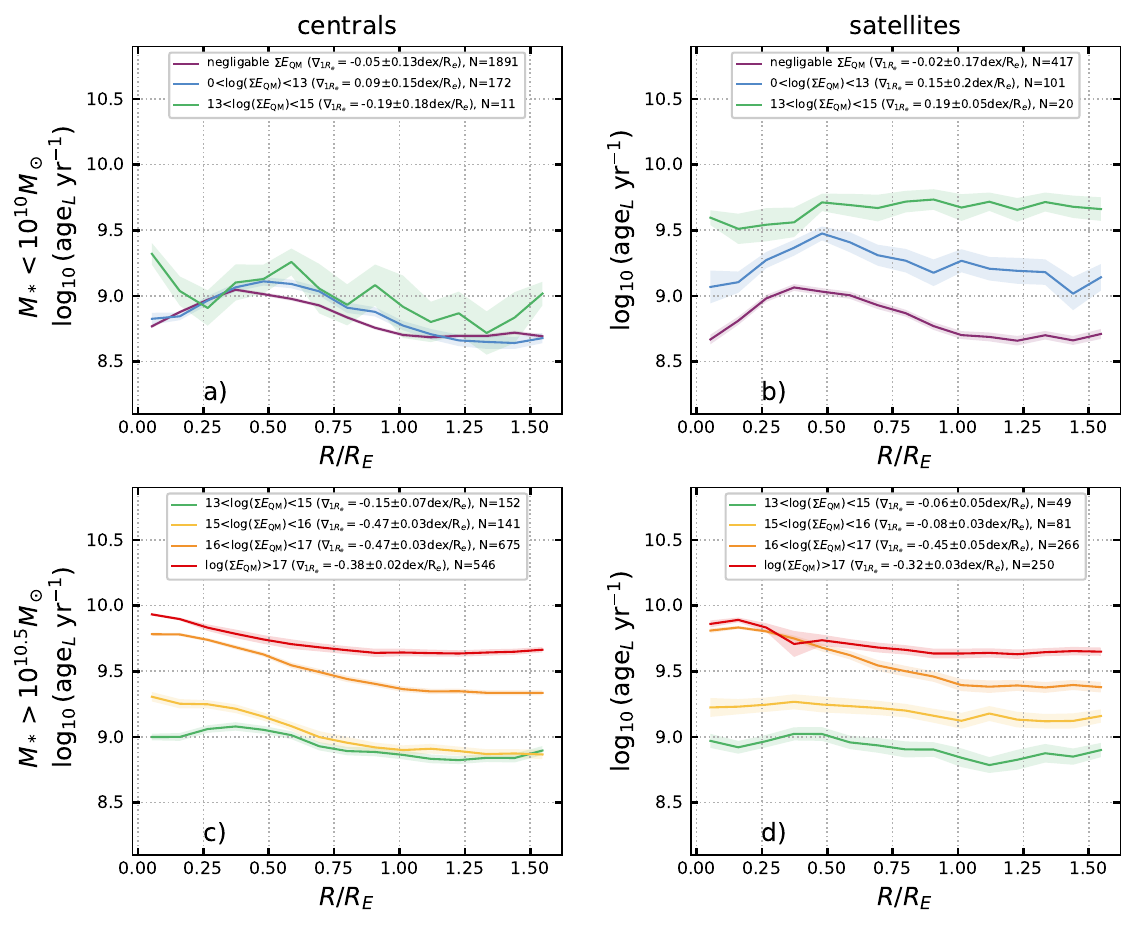}
    \caption{Radial profiles of \age{}, subdivided by cumulative AGN feedback energy imparted in the quasar mode, $\Sigma E_{\mathrm{QM}}= \Sigma (E_{\mathrm{QM}}/(M_{\odot}\mathrm{kpc}^2 \mathrm{Gyr}^{-2}))$, where $E_\mathrm{QM}$ has units of $M_{\odot}\mathrm{kpc}^2 \mathrm{Gyr}^{-2}$. Formatting is identical to Figure \ref{fig:BHmass}.}
    %The legends contain computed profile gradients, and $N$, the number of galaxies that contributed to the profile. Profiles are computed separately for populations of high-mass (top row) and low- mass (bottom row) central (left column) and satellite  (right column) galaxies. Errors are computed from 2000 bootstrap resamplings. Error bounds that are wider than the linewidth are indicated by shaded regions and are omitted from the figure when they are comparable to or smaller than the linewidth.
    
    \label{fig:cumQM}
  
\end{figure}

\begin{figure}
    \centering
    \includegraphics[width=\textwidth, keepaspectratio]{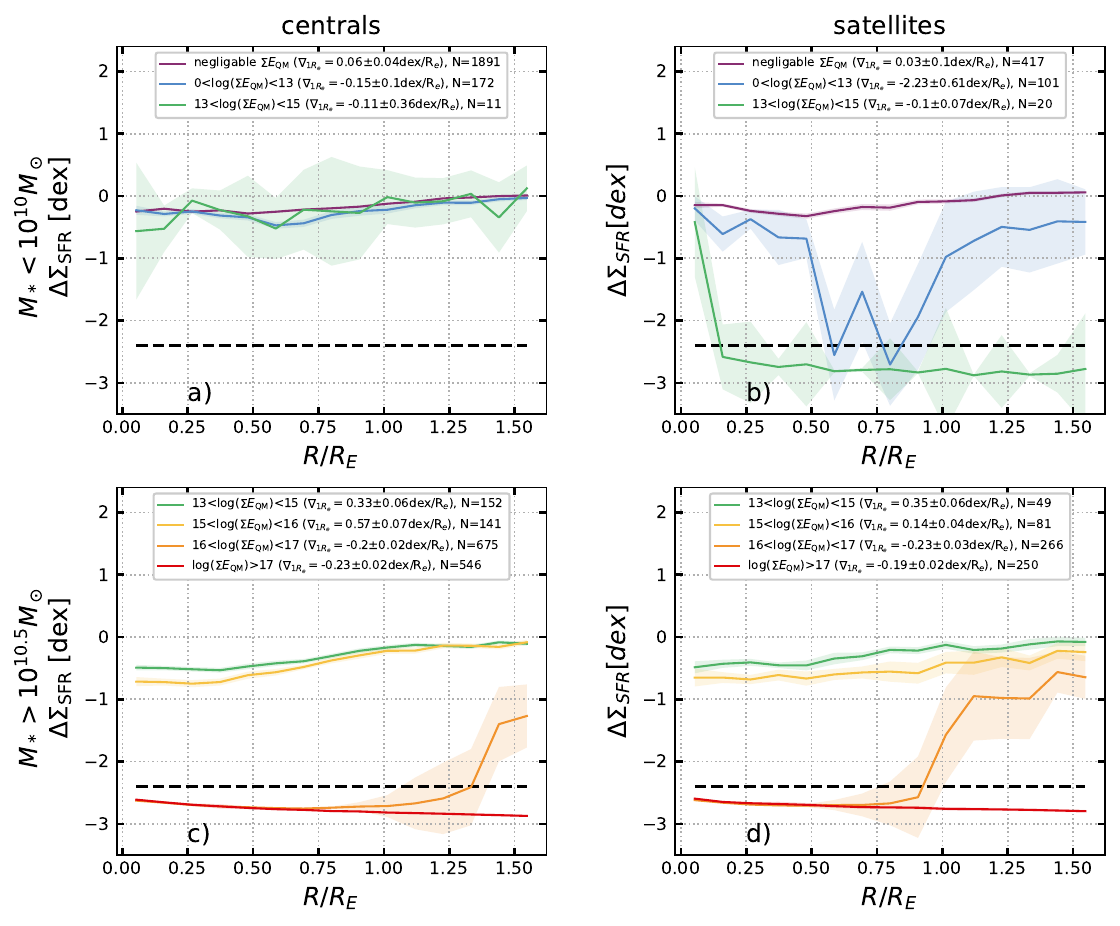}
    \caption{\textnormal{Radial profiles of \SFR{}, subdivided by cumulative AGN feedback energy imparted in the quasar mode, and corresponding to Figure \ref{fig:cumQM}. Here, the formatting is identical to that of Figure \ref{fig:SFR_BHMass}}.}
    \label{fig:SFR_cumQM}
  
\end{figure}

In Figures \ref{fig:cumQM}\textnormal{ and \ref{fig:SFR_cumQM}}, the formatting is identical to Figure \ref{fig:BHmass} and \ref{fig:SFR_BHMass}, except profiles are subdivided by the cumulative energy imparted by the SMBH in the quasar mode, \cumQM. The units of \cumQM{} that we use are the reduced ``code units'' of $M_\odot {\rm kpc^2 Gyr^{-2}}$.  For sake of brevity within the text, we do not explicitly include the units of \cumQM{} below.

Figures \ref{fig:cumQM}\textnormal{ and \ref{fig:SFR_cumQM} show} only profiles for galaxies with seeded SMBHs, although some of these black holes have imparted negligible feedback energy. For the high-mass galaxy populations in panels c) and d), both the shape and normalization of the profiles vary with different values of \cumQM. For both high-mass central and satellite galaxies, the normalization of the \age{} profiles increases with increasing \cumQM. Notably, the steepest \age{} profiles for high-mass galaxies are not found in the bin with the highest \cumQM, rather they are found in the intermediate bins. 

\textnormal{In Figure \ref{fig:SFR_cumQM}, high-mass galaxies with the greatest \cumQM{} have profiles that lie entirely below the quenching threshold. While the centers of galaxies with $10^{16}<$\cumQM$<10^{17}$ (orange) are completely quenched, there is still ongoing star formation in the outskirts (at $\gtrsim1.25 R_e$ for centrals and $\gtrsim0.9 R_e$ for satellites. }
%This may indicate that high-mass galaxies with \cumQM$>10^{17}$ may already be completely quenched, while central galaxies with $10^{15}<$\cumQM $<10^{17}$ and satellite galaxies with $10^{16}<$\cumQM $<10^{17}$ are still undergoing quenching driven by the SMBH feedback energy. 

Moving to the low-mass population, we find that \cumQM{} has little effect on the profiles of central galaxies.  This, however, is not the case for satellite galaxies, where we find that the normalization of \age{} profiles for low-mass satellite galaxies increases with increasing \cumQM. \textnormal{The population of low-mass satellite galaxies with $10^{13}<$\cumQM$<10^{15}$ (green) have averaged \SFR{} profiles that are almost entirely quenched, except at the very center. } These results are somewhat surprising, since \cumQM{} is correlated with $M_{\mathrm{BH}}$, and the results shown in Figures \ref{fig:BHmass} \textnormal{and \ref{fig:SFR_BHMass}} indicate no difference in profile normalization as a function $M_{\mathrm{BH}}$ for the low-mass satellite population. 
%This, combined with the lack of this trend in low-mass centrals, could indicate that the same process that is driving up the \age{} normalization is also driving the increase in \cumQM{} for the low-mass satellite population. However, %It does not appear that \cumQM{} is directly driving quenching in low-mass satellites, due to the flat slopes of the profiles.

%There is no significant trend in profile shape or normalization for \age{} profiles of galaxies separated by the cumulative energy imparted in the low-accretion state radio mode, $\Sigma E_{RM}$, not shown here. There are less galaxies with any energy imparted in this state, which generally only turns on after a galaxy has exhausted its supply of star-forming gas and is considered quenched. 
There are only $337$ galaxies in our sample with non-negligible AGN fedback energy imparted in the radio mode, $\Sigma E_{RM}$. This feedback mode only turns on at low accretion rates, typically in massive, quenched galaxies \citep{Weinberger18}. The low number of galaxies with any $\Sigma E_{RM}$ makes it impossible to draw firm conclusions about sub-populations. 

Profiles of populations subdivided by total AGN feedback energy imparted in both states (not shown) appear very similar to Figures \ref{fig:cumQM} \textnormal{and \ref{fig:SFR_cumQM}}, although there is no longer a separation in the normalization of low-mass satellite populations. Instead, the two bins with non-negligible feedback energy both appear similar in shape and normalization to the profile for low-mass satellites with $10^{0}<\Sigma E_{QM}<10^{13}$ \textnormal{(blue)}. When looking at total feedback energy for low-mass satellites, the bin with $10^{13}<\Sigma E_{AGN}<10^{15}$ \textnormal{(green)} contains two distinct populations: those that injected energy primarily in the radio (kinetic) mode and those that have operated primarily in the quasar (thermal) mode. Combining these populations results in a lower overall normalization than seen in the $10^{13}< \Sigma E_{QM}<10^{15}$ \textnormal{(green)} low-mass satellite population.
%this population obfuscates the results for low-mass satellites with $13<\log(\Sigma E_{\mathrm{AGN}})<15$, as the profile combines populations of satellites with SMBHs that have operated primarily in the quasar mode and those that have operated primarily in the radio mode. Otherwise, there is no significant difference between the profiles of galaxy populations separated by $\Sigma E_{\mathrm{AGN}}$ and \cumQM. 

\subsection{Morphology} \label{sec:morph}

\begin{figure}
    \centering
    \includegraphics[width=\linewidth]{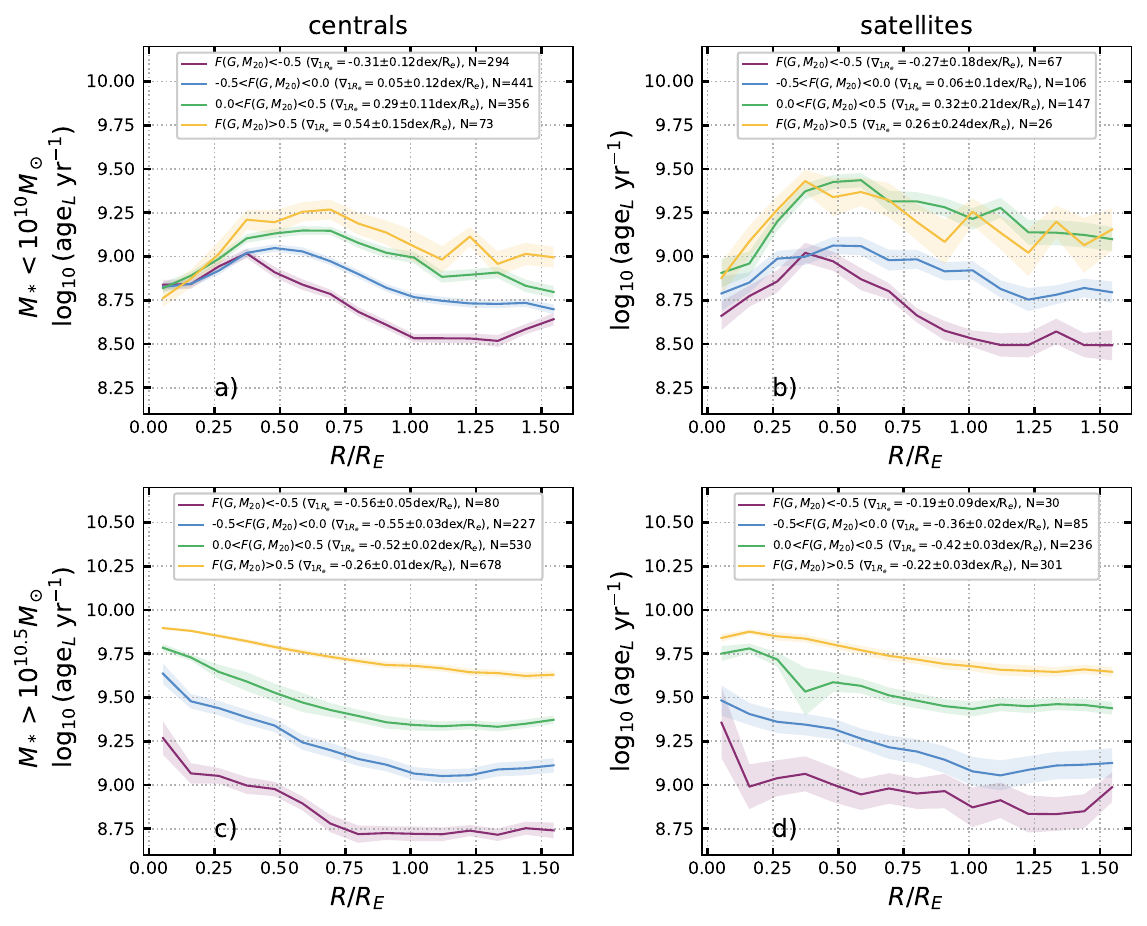}
    \caption{Radial profiles of \age{} subdivided by the strength of their bulge as measured by the Gini-M20 bulge statistic, \bulge. Positive values of \bulge{} correspond to bulge-like morphology, with higher values indicating stronger bulges. Formatting is identical to Figure \ref{fig:BHmass}.}

    \label{fig:bulge}
  
\end{figure}

\begin{figure}
    \centering
    \includegraphics[width=\textwidth, keepaspectratio]{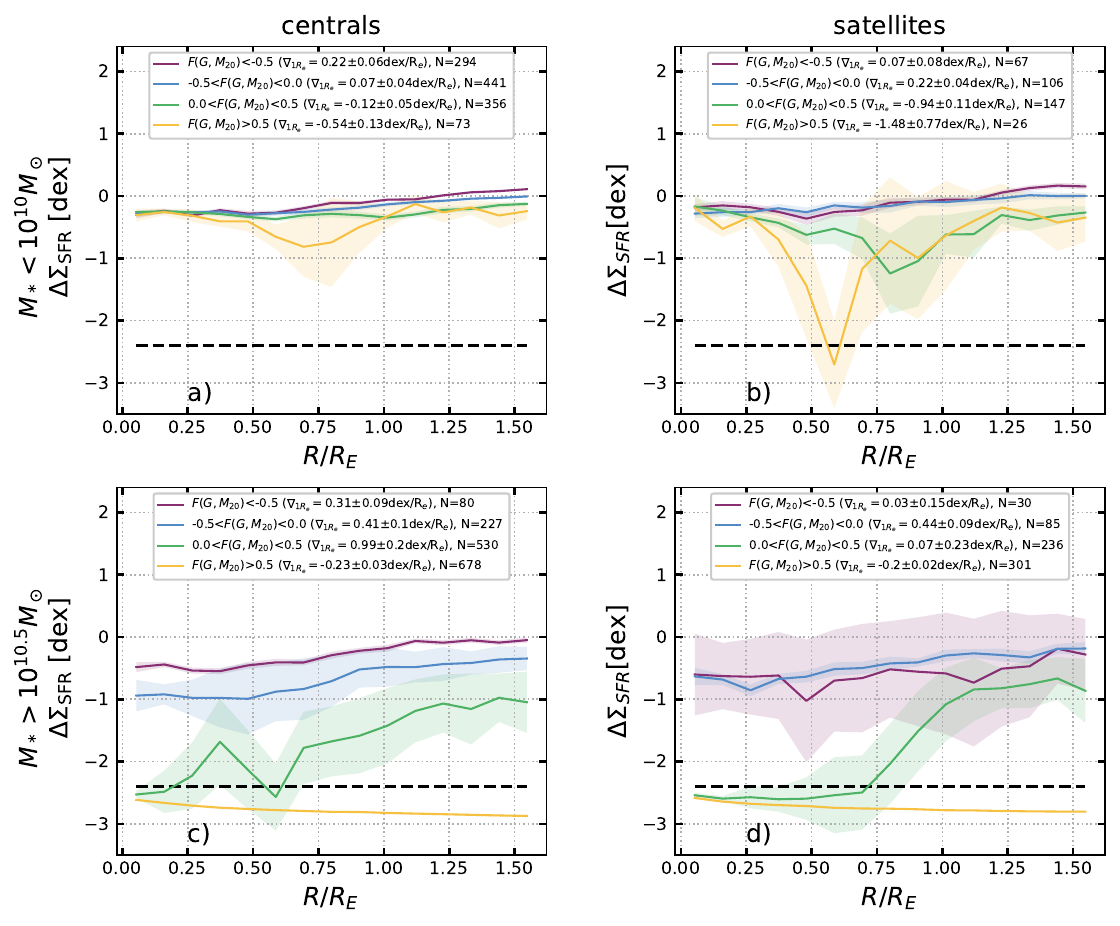}
    \caption{\textnormal{Radial profiles of \SFR{}, subdivided by the Gini-M20 bulge statistic, \bulge, and corresponding to Figure \ref{fig:bulge}. Here the formatting is identical to that of Figure \ref{fig:SFR_BHMass}.}}
    \label{fig:SFR_bulge}
  
\end{figure}

Some studies \citep[e.g.,][]{Martig}, have indicated that the presence of a galactic bulge can \textnormal{decrease star formation efficiency by stabilizing gas and thereby regulating} star formation. In Figures \ref{fig:bulge} \textnormal{and \ref{fig:SFR_bulge}}, we present radial profiles of \age{} \textnormal{and \SFR{}} for populations of galaxies binned by the Gini-M20 bulge statistic, \bulge. A positive \bulge{} value indicates the presence of a bulge, with higher values corresponding to stronger bulges. For different populations of galaxies in Figure \ref{fig:bulge}, there is a difference in \textnormal{\age{}} profile shape and normalization between galaxies with bulge-like morphologies (\textnormal{green and yellow}) and those without (\textnormal{purple and blue}). Galaxies with bulges have older ages throughout, with the exception of low-mass central galaxies at small radii.

For high-mass galaxies (bottom row, Figure \ref{fig:bulge}), the \age{} profiles increase in normalization with increasing \bulge, but the slopes for all populations except the most bulge-like agree within $1\sigma$ for centrals and $2\sigma$ for satellites. Galaxies with \bulge$>0.5$ \textnormal{(yellow)} have shallower \age{} profile gradient than galaxies with weaker bulges. \textnormal{In the corresponding \SFR{} profiles, Figure \ref{fig:SFR_bulge} shows that the \bulge$>0.5$ (yellow) population is completely quenched, while the $0<$\bulge$<0.5$ (green) population has a \SFR{} profile characteristic of green valley galaxies.} We note that, as shown in Appendix~\ref{app_corr}, the bulge parameter is well-correlated with $M_{\rm BH}$ and \cumQM.

For the low-mass galaxies (top row, Figure \ref{fig:bulge}), there is a difference in the normalization of the \age{} profiles and, to a lesser degree, the shapes of the profiles. Bulges tend to contain older stellar populations \citep{Sarzi05}, so the effect on the \age{} profiles may not reflect changes to star formation rates. \textnormal{Indeed, the \SFR{} profiles in Figure \ref{fig:SFR_bulge} show little dependence on \bulge. }

\textnormal{At intermediate radii, \SFR{} profiles of satellites with bulges (\bulge$>0$) are being overwhelmed by spaxels that lack star formation, which is a known limitation in our analysis (see \citealt{McDonough23}). In essence, there is a minimum resolvable star formation rate in TNG100. Resolution also limits detectable star formation levels in observations. To address this, \cite{Bluck2020b} used a binning procedure that boosted the signal-to-noise, resulting in more low-SFR than high-SFR spaxels being binned together. In our analysis, we do not perform a similar binning procedure. The presence of low-\SFR{} features in the top panels of Figure \ref{fig:SFR_bulge} do indicate that there is a greater percentage of quenched spaxels in those regions of bulge-like TNG100 galaxies, but we cannot draw firm conclusions about the shapes of the profiles. }

We find that the \age{} profiles for galaxies subdivided by \sersic{} indices are remarkably similar to those for the bulge statistic and we opt not to present them here.  Instead, we focus on results as a function of \bulge{} since this parameter is better at identifying bulges in satellites that may have been disturbed by interactions \citep{Snyder2015}.

\section{Dependence on Environmental Parameters} \label{sec:envprof}
\subsection{Halo mass}

\begin{figure}
    \centering
    \includegraphics[width=\linewidth]{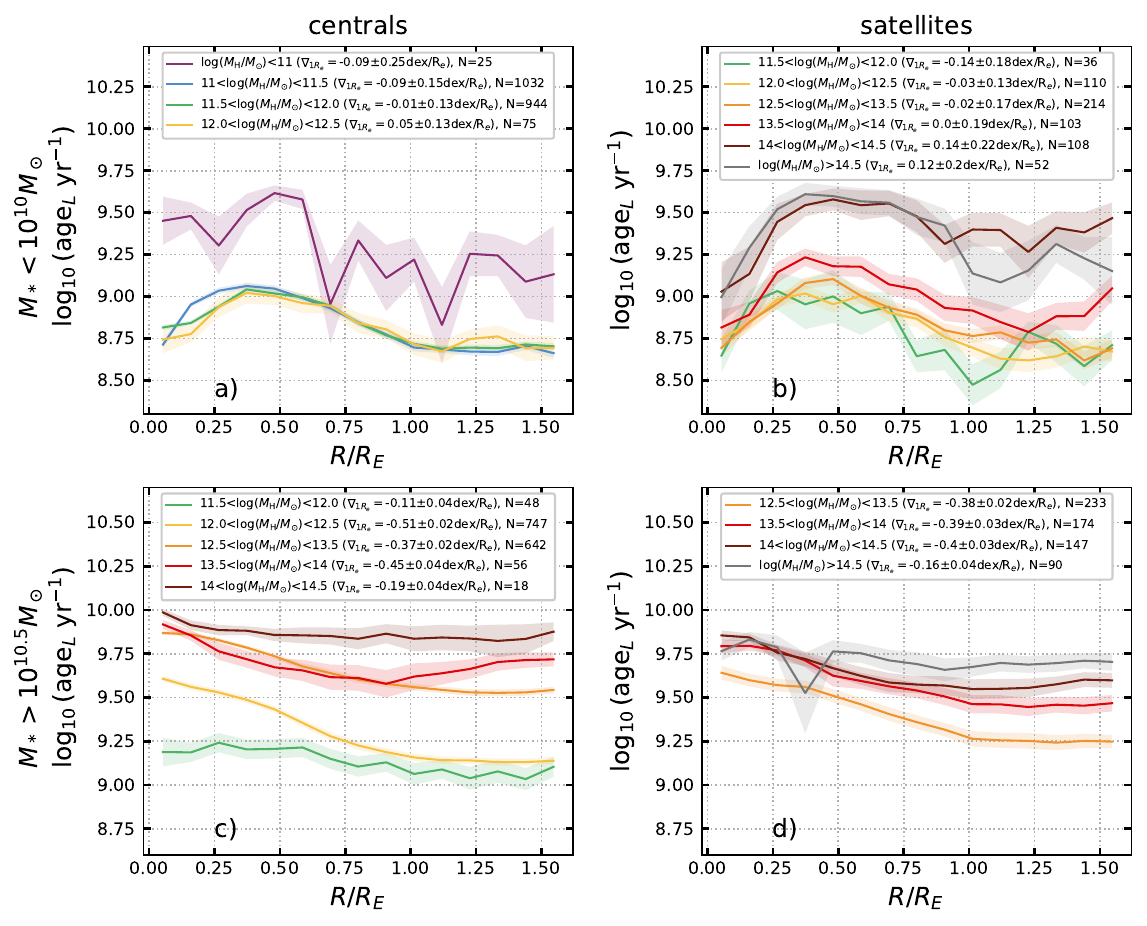}
    \caption{Radial profiles of \age{} subdivided by halo mass, $M_{H}$. Formatting is identical to Figure \ref{fig:BHmass}.}
    \label{fig:halo}
  
\end{figure}

\begin{figure}
    \centering
    \includegraphics[width=\textwidth, keepaspectratio]{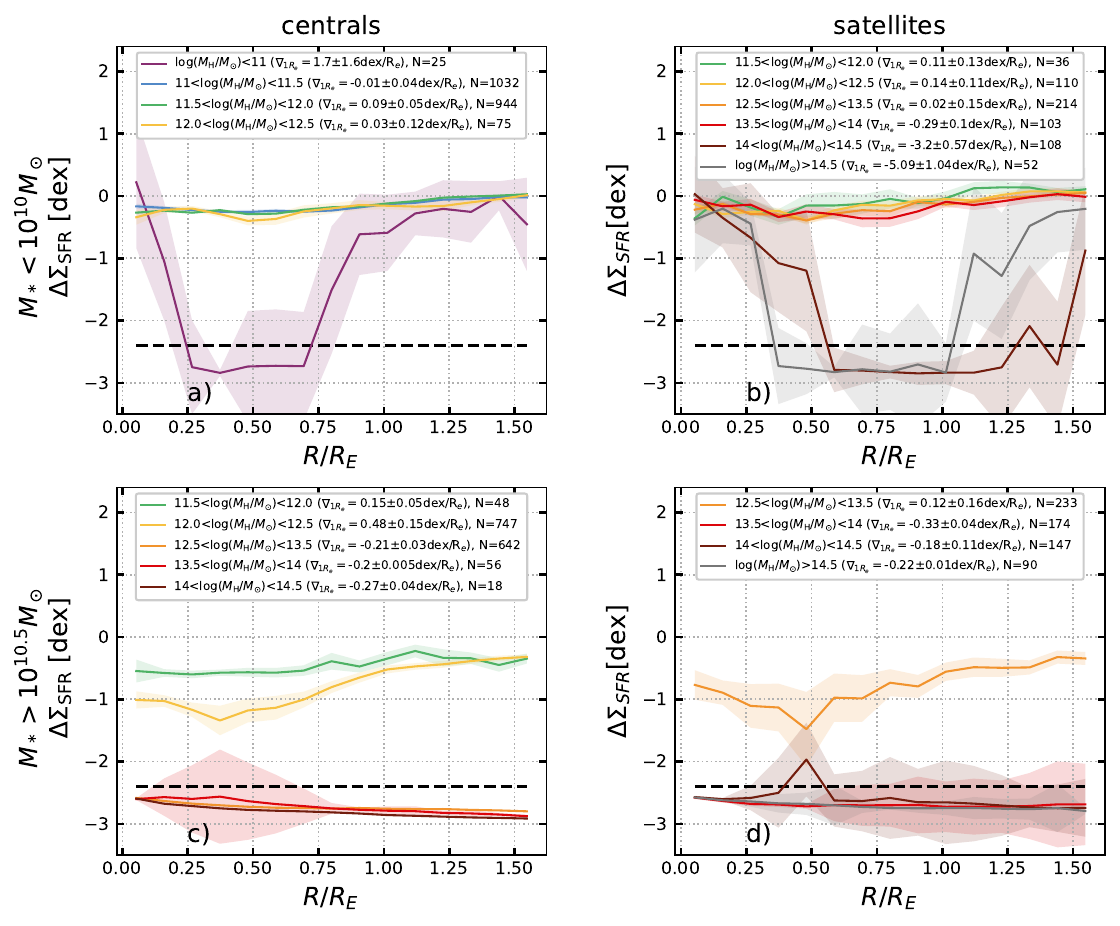}
    \caption{\textnormal{Radial profiles of \SFR{}, subdivided by host halo mass, and corresponding to Figure \ref{fig:halo}. Here the formatting is identical to Figure \ref{fig:SFR_BHMass}.}}
    \label{fig:SFR_halo}
  
\end{figure}

In Figures \ref{fig:halo} \textnormal{and \ref{fig:SFR_halo}} we present the dependence of the distribution of \age{} \textnormal{and \SFR{}} on halo mass. For the central galaxy population, halo mass scales with SMBH mass and, thus, \cumQM{} (see Appendix \ref{app_corr}). As in \textnormal{\S \ref{sec:SMBH}}, there is little difference in normalization or shape in the profiles of the low-mass central population when subdivided by halo mass, except at the very lowest masses ($M_H/M_\odot <10^{11}$\textnormal{, purple}). For high-mass centrals, all populations have negative \textnormal{\age{}} gradients that are indicative of inside-out quenching, although the gradient is significantly shallower for galaxies in halos with $M_H/M_\odot <10^{12}$ and $M_H/M_\odot >10^{14}$. The slope of the \textnormal{\age{}} profile of central galaxies in halos with $10^{13.5}< M_{H}<10^{14}$ \textnormal{(red)} becomes positive at large radii ($\gtrsim 1R_e$), deviating from the flat outskirts seen in the other high-mass central populations.
\textnormal{In Figure \ref{fig:SFR_halo}, the high-mass central population with $M_H/M_\odot >10^{12.5}$ have average \SFR{} profiles that lie entirely below the quenching threshold.}

\textnormal{In Figure \ref{fig:halo}, }both high- and low-mass satellite galaxies that reside in high-mass halos ($M_{H}>10^{13.5}$) have older overall \age{} profile normalizations than do satellites residing in halos with $M_{H}<10^{13.5}$. The difference is much larger for low-mass satellites than it is for high-mass satellites, especially at $M_H/M_\odot>10^{14}$. At radii $>1.25 R_e$, the slopes of the profiles for low-mass satellites with $10^{13.5}<M_H/M_\odot <10^{14}$ \textnormal{(red)} and $10^{14}<M_H/M_\odot <10^{14.5}$ \textnormal{(brown)} become positive. Similar to the high-mass centrals, the \age{} profiles for high-mass satellites become flatter at high halo masses, although this occurs at higher halo masses for the satellites.

\textnormal{The low-mass satellite galaxy populations with high halo masses ($M_H/M_\odot >10^{14}$ in Figure~\ref{fig:SFR_halo}b) are highly variable and suffer from the issues with low-\SFR{} profiles discussed above and in \cite{McDonough23}. We caution against over-interpreting these profiles, however we can conclude that this population has a higher percentage of quenched spaxels than populations at lower halo masses. The high-mass satellites in Figure~\ref{fig:SFR_halo}d) are also highly variable, resulting in large error bars. On average, high-mass satellites in halos with $M_H/M_\odot>10^{13.5}$ are almost entirely quenched.}

\subsection{Local Galaxy Overdensity}
\begin{figure}
    \centering
    \includegraphics[width=\linewidth]{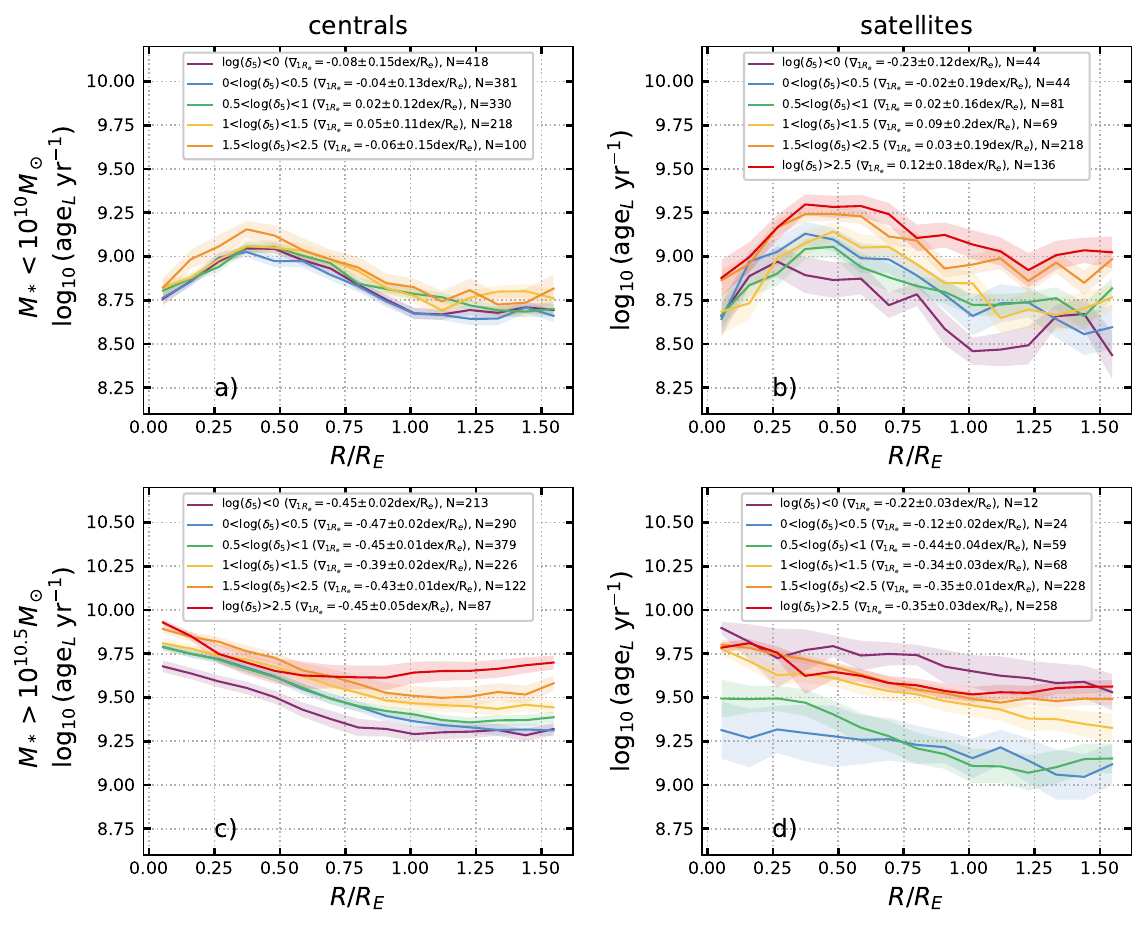}
    \caption{Radial profiles of \age{} subdivided by local galaxy overdensity (see text). Formatting is identical to Figure \ref{fig:BHmass}.}
    \label{fig:overdens}
  
\end{figure}

\begin{figure}
    \centering
    \includegraphics[width=\textwidth, keepaspectratio]{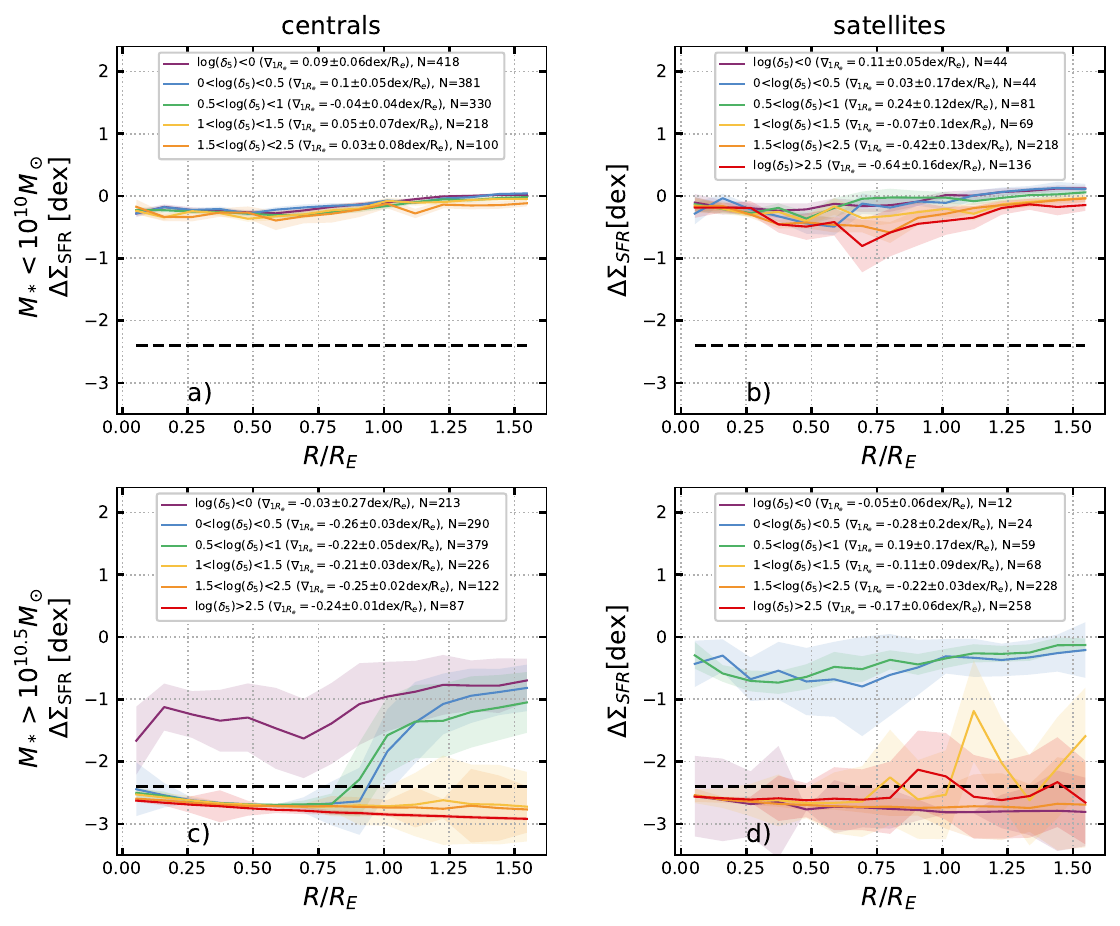}
    \caption{\textnormal{Radial profiles of \SFR{}, subdivided by local galaxy overdensity measured at the fifth nearest neighbor, that correspond to Figure \ref{fig:halo}. Here the formatting is identical to that of Figure \ref{fig:SFR_BHMass}.}}
    \label{fig:SFR_overdens}
  
\end{figure}

Figures~\ref{fig:overdens} \textnormal{and \ref{fig:SFR_overdens}} present \textnormal{\age{} and \SFR{}} profiles of TNG100 galaxies subdivided into populations based on local galaxy overdensity measured at the $5^{\rm th}$ nearest neighbor. \textnormal{For low-mass central galaxy populations, there is very little deviation in the profiles of \age{} and \SFR{} over a large range of $\delta_5$.} In the case of high-mass central galaxies, the \age{} profiles show slight differentiation in the normalization as a function of $\delta_5$, although not to the same degree as we found above for halo mass. However, the shapes of \age{} profiles for high-mass galaxies with $10^{1.5}< \delta_5 <10^{2.5}$ \textnormal{(orange)} have a slightly positive slope in the average profile at large radii. The slope becomes more positive in the outskirts of the high-mass central population in even denser regions ($\delta_5 > 10^{2.5}$,\textnormal{ red}). Compared to other populations, high-mass central galaxies in very underdense regions ($\delta_5 <10^{0}$, \textnormal{purple}) have younger ages at $R_e<1.0$ and they are the only high-mass central population that is still star forming throughout \textnormal{in Figure~\ref{fig:SFR_overdens}c)}. 

In contrast, the few satellite galaxies that are found in the most underdense regions have the oldest ages\textnormal{ and are quenched throughout}, comparable to satellites in the most overdense regions. Aside from high-mass satellites in very underdense regions, there is a weak trend of increasing normalization of \age{} with increasing $\delta_5$ for satellites. \textnormal{This trend is weaker in the \SFR{} profiles, although low-mass satellites in especially overdense regions have somewhat lower SFRs at some radii.} The widths of the error bars on profiles of populations subdivided by $\delta_5$ are generally larger than those of populations subdivided by $M_{\rm H}$, indicating greater variability in the profiles of individual satellites in the $\delta_5$ populations.

\subsection{Satellite Parameters}

\subsubsection{Joining time}
The environmental processes that affect star formation in satellite galaxies are expected to scale with time spent in the host halo. More time spent in the halo will lead to a higher probability of interactions and more time for gas to be stripped from the satellite.

\begin{figure}
    \centering
    \includegraphics{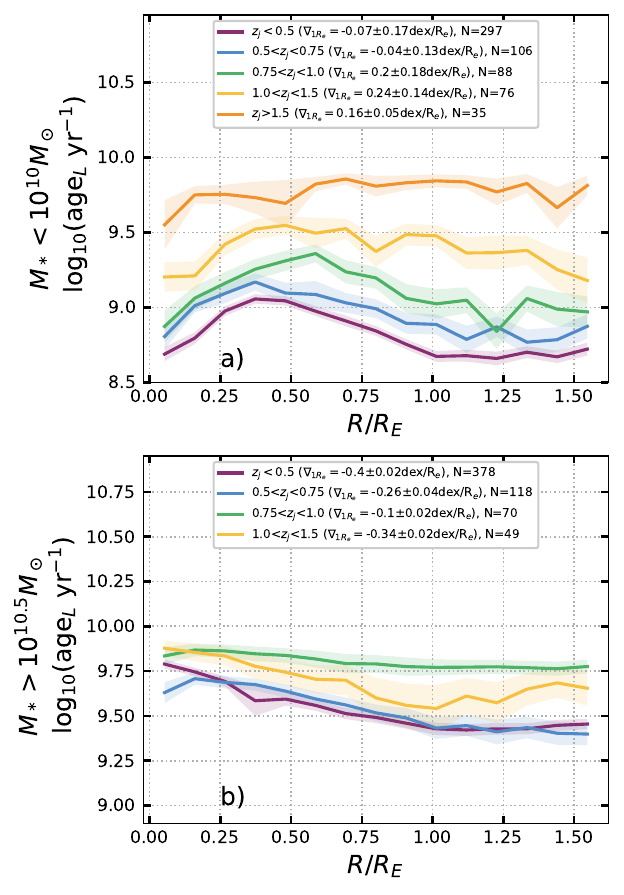}
    \caption{Radial profiles of \age{} subdivided by the redshift at which the satellite entered within $3R_{200}$ of its $z=0$ host, $z_j$. Formatting is identical to the right column of Figure \ref{fig:BHmass}.}
    \label{fig:joinz}
  
\end{figure}

\begin{figure}
    \centering
    \includegraphics{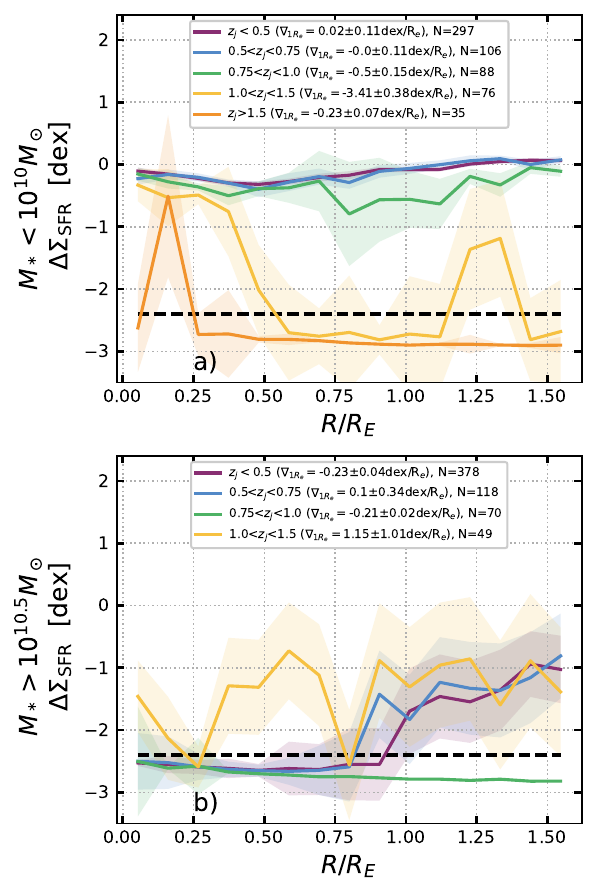}
    \caption{\textnormal{Radial profiles of \SFR{}, subdivided by the redshift at which the satellite approached within $3R_{200}$ of its $z=0$ host, $z_j$, and corresponding to Figure \ref{fig:joinz}. Here the formatting is identical to the right column of Figure \ref{fig:SFR_BHMass}.}}
    \label{fig:SFR_joinz}
\end{figure}

In Figures~\ref{fig:joinz} \textnormal{and \ref{fig:SFR_joinz}}, we present radial profiles of \age{} \textnormal{and \SFR{}} for satellites, with the sample subdivided based on the redshift at which the satellites joined their hosts, $z_j$. The low-mass population shows that the overall \age{} increases for satellite populations with greater $z_j$. \textnormal{Low-mass satellites with $z_j>1.0$ have lower \SFR{} profiles than satellites that joined later. On average, low-mass satellites with $z_j>1.5$ only maintain some star formation in their centers. }
%This is true to a lesser degree for the high-mass satellite population.  

For the high-mass galaxies, there is a curious reversal in overall age between satellites in the $0.75< z_j <1.0$ \textnormal{(green)} and $1.0< z_j <1.5$ \textnormal{(yellow)} populations, with the population that joined earlier having a younger overall \age. \textnormal{This is also reflected in the \SFR{} profiles, with the $1.0< z_j <1.5$ population maintaining some degree of star formation while the $0.75< z_j <1.0$ population appears entirely quenched. High-mass satellites with relatively late joining times ($z_j<0.75$) have the characteristic inside-out \age{} and \SFR{} profile shapes.}

Compared to more recently accreted low-mass satellites, low-mass satellites with $z_j >1.0$ have \age{} profiles that are flatter beyond $R>0.75 R_e$. High-mass satellites with $z_j >0.75$ have flat \age{} profiles. The shapes and normalizations of these profiles indicate that more outside-in quenching has operated on satellite populations that have spent longer in their hosts' environment, as expected. However, there is a high degree of variation within a given population, as indicated by the shaded error bars. 

\subsubsection{Change in mass of star forming gas}

\begin{figure}
    \centering
    \includegraphics{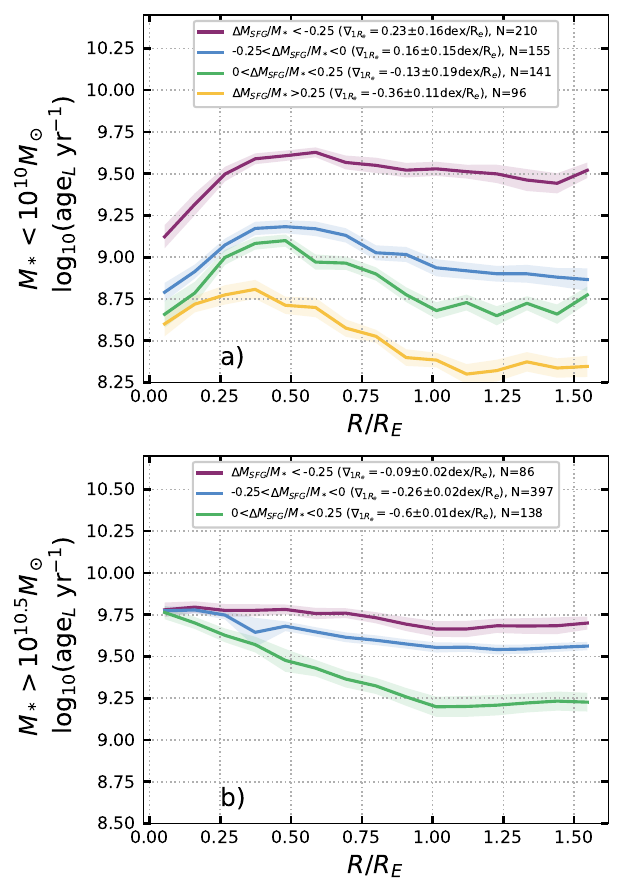}
    \caption{Radial profiles of \age{} subdivided by the change in star-forming gas since the satellite entered within $3R_{200}$ of its $z=0$ host, \sfgas. Formatting is identical to Figure \ref{fig:joinz}. Note that the ordinates of each panel have slightly different numerical ranges.}
    \label{fig:SFgas}
  
\end{figure}

\begin{figure}
    \centering
    \includegraphics{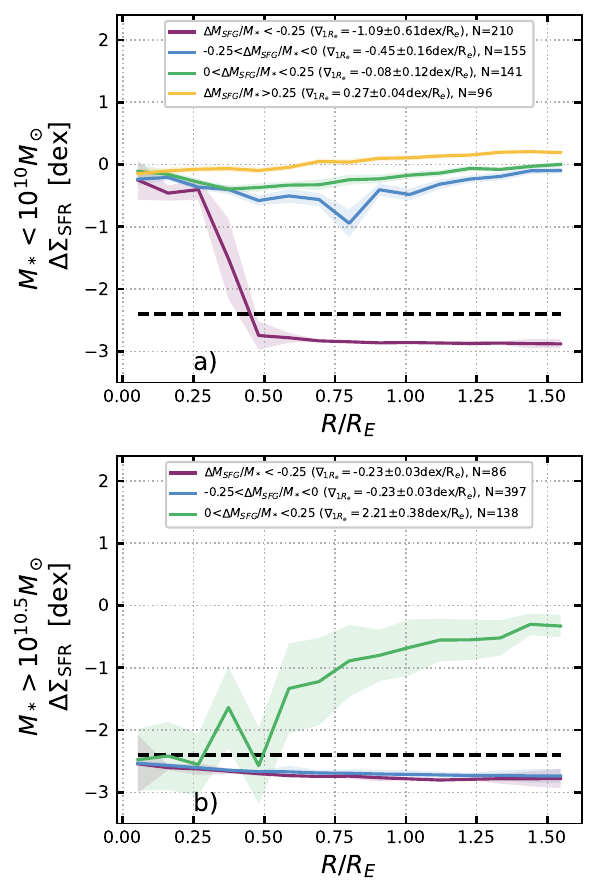}
    \caption{\textnormal{Radial profiles of \SFR{}, subdivided by the change in star-forming gas since the satellite approached within $3R_{200}$ of its $z=0$ host, and corresponding to Figure \ref{fig:joinz}. Here the formatting is identical to the right column of Figure \ref{fig:SFR_BHMass}.}}
    \label{fig:SFR_SFgas}
\end{figure}

Any process that affects star formation in galaxies affects the availability of star-forming gas (SFG). In the simulation, it is possible to determine the change in star-forming gas since a satellite joined the host halo, \sfgas. In Figures \ref{fig:SFgas} \textnormal{and \ref{fig:SFR_SFgas}}, we present radial profiles of \age{}\textnormal{ and \SFR{}} subdivided by \sfgas, and normalized by the stellar mass of each satellite. For low- and high-mass satellites, there is a clear offset in overall age \textnormal{and \SFR{}} for different bins of \sfgas. Satellites that, proportionately, have lost more star-forming gas are systematically older\textnormal{ and have lower \SFR{}}, while those that have gained star-forming gas are systematically younger\textnormal{ and have higher \SFR{}}. For low-mass galaxies, the difference in \age{} for satellites with \sfgas$/M_* <-0.25$ and satellites with \sfgas$/M_* >0.25$ is nearly $3$ Gyr ($\sim 1$ dex). \textnormal{While low-mass satellites with significant SFG loss maintain typical levels of star formation at their centers ($R<0.25R_e$), at radii $>0.5 R_e$ they are completely quenched.}

There are no high-mass satellite galaxies that have gained more than $25\%$ of their stellar mass in star-forming gas. The age difference between high-mass satellites with \sfgas$/M_* <-0.25$ \textnormal{(purple)} and high-mass satellites that have gained any star-forming gas is about $3.5$ Gyr ($\sim0.5$ dex) in the outskirts. \textnormal{The high-mass satellites with positive \sfgas{} }(green)\textnormal{ are the only population of high-mass satellites that show significant levels of star formation, which is occurring at radii $>0.5R_e$. }

There is a clear difference in \age{} profile shape for satellites that have gained star-forming gas and those that have lost star-forming gas. Satellites with negative \sfgas{} are older than satellites with positive \sfgas{}, particularly in the outskirts ($R \gtrsim 0.5 R_e$). \textnormal{These results demonstrate} that the change in the availability of star-forming gas is occurring in the outskirts of these simulated galaxies, consistent with outside-in quenching or enhancement. %This is confirmed by the profiles of \SFR{} in Figure~\ref{fig:SFR_SFgas}. Low-mass satellite galaxies that have lost the most star-forming gas are the only low-mass population that is quenched in any region ($R \gtrsim 0.5 R_e$). High-mass satellite galaxies that have gained star-forming gas are the only high-mass population that are still actively star-forming in any region. 
%In both cases, the quenching or enhancement in star formation occurs at $R \gtrsim 0.5 R_e$. 

\subsubsection{Perigalacticon}
For each satellite in our sample, we obtained the redshift at which it last made its closest approach, $z_{\mathrm{p}}$, as well as the closest approach the satellite made to its host, $d_{h-s}$. Here, we briefly summarize \textnormal{the} \age{} \textnormal{and \SFR{}} profiles of galaxy populations subdivided by these parameters, although we refrain from showing them since, for the most part, the differences are relatively minor.

There is no significant dependence of \age{} profile on $d_{h-s}$ for high-mass satellite galaxies. For low-mass satellites, significant differences only appear for satellites with $d_{h-s}<0.5R_{200}$. The outskirts of low-mass satellites with $d_{h-s}<0.5R_{200}$ are $\sim0.4$ dex ($\sim550$ Myr) older than satellites with $d_{h-s}>2R_{200}$. There is no significant trend in outskirt ages when finer bin sizes are used to define populations at smaller host-satellite separations. \textnormal{The corresponding \SFR{} profiles are very noisy, indicating a high degree of variability among the individual galaxy profiles. Profiles of low-mass satellites are nearly identical. There is no significant trend with $d_{h-s}$ for high-mass galaxies, with most profiles appearing entirely quenched or indicative of typical inside-out quenching. The only exception is the profile of satellite galaxies with the smallest perigalacticon separation, $d_{h-s}<0.1 R_{200}$. This profile is quenched in the center and outskirts, but maintains active star formation at $0.5\lesssim R/R_e \lesssim 1.2$. However, this is one of the noisiest profiles, with errors in the range of $\sigma_{\Delta \Sigma_{\rm SFR}}\approx 0.75-0.95$ at $R/R_e>0.5$.  }

For $z_{\mathrm{p}}$, the \age{} profiles of low-mass satellites have a normalization trend that is similar to that of populations subdivided by $z_j$ (Figure \ref{fig:joinz}), with satellites with larger values of $z_{\mathrm{p}}$ having older ages. Satellites with $z_{\mathrm{p}}>0.5$ have the oldest ages and flatter \age{} profiles. 
%These satellites also tend to have earlier joining redshifts. 
The trend in increasing \age{} normalization with increasing $z_{\mathrm{p}}$ is weaker for high-mass satellites than it is for low-mass satellites. Satellites with $z_{\mathrm{p}}>0.1$ have ages that are $\sim0.2$ dex ($\sim1.8$ Gyr) older than satellites with the latest perigalacticon approaches. This is a significantly smaller difference than satellites with the earliest and latest $z_j$, which is $\sim0.5$ dex ($\sim5$ Gyr).

\textnormal{The \SFR{} profiles subdivided by $z_{\rm p}$ are very noisy. The profiles correspond well with insights gleaned from the complimentary \age{} profiles. Low-mass satellites with $z_{\rm p}>0.5$ are the only low-mass satellite population with a profile that lies below the quenching threshold at any radius ($R/R_e >0.3$). The high-mass satellite profiles are indicative of active inside-out quenching, with the exception of the population with $z_{\rm p}>0.1$, which is entirely quenched. }

\textnormal{It is not surprising that satellite populations with the earliest perigalacticon approaches are quenched, as these populations also have the earliest joining times. However, the reverse is not necessarily true, as satellites with early joining times may have relatively recent perigalacticon approaches. The noisiness of the \age{} and \SFR{} profiles, along with the relatively weak trends with $d_{h-s}$ and $z_{\rm p}$, lead us to believe that these parameters play a relatively minor, if any, role in quenching satellites.}

\section{Discussion} \label{sec:discussion}
In this paper, we have considered various parameters that are related to the two main quenching processes identified in the literature: mass quenching and environmental quenching.

In the mass quenching scenario \citep[see, e.g.,][]{Peng10,Ma22,Bluck2020b}, energy from the central supermassive black hole \textnormal{acts on} gas from the interstellar and circumgalactic media, preventing further accretion and star formation. \textnormal{The exact mechanism by which the SMBH affects star formation is still unclear, but it may be a combination of heating, removal, or the introduction of turbulance that decreases star formation efficiency.} The \textnormal{degree to which AGN feedback negatively affects star formation in a galaxy} increases with increasing amounts of energy imparted by the black hole. This process is known as mass quenching because the energy scales with black hole mass and, thus, stellar mass. Mass quenching operates at stellar masses greater than $\sim 10^{10} M_{\odot}$ \citep[e.g.,][]{Bluck2020b, McDonough23}. 

 \textnormal{Intrinsic feedback processes in low-mass galaxies are not powerful enough to quench these systems, but do act to slow the natural exhaustion of their gas reservoirs, which generally allows these systems to maintain star formation at present-day.} Other processes must be necessary to quench galaxies at low masses, since quenched galaxies at $M_*\lessapprox10^{10} M_{\odot}$ do exist. These low-mass, quenched galaxies consist primarily of satellite galaxies that reside in group and cluster environments, where environmental processes are common. Environmental quenching is any external process that negatively affects the availability of star forming gas in a galaxy, and is particularly important for satellites \citep[see, e.g.,][]{Peng12,Bluck2020a,Bluck2020b, Corcho-Cabellero2023}.

As mass quenching is the dominant process in high-mass galaxies, defined as \himass{} in this paper, we will discuss our results for high-mass and low-mass (\lomass{}) galaxies separately.

\subsection{High-mass galaxies}
In Figures \ref{fig:BHmass} and \ref{fig:cumQM} (bottom panels), mass quenching is clearly operating in both high-mass central and satellite galaxies. More massive black holes and higher values of \cumQM{} \textnormal{correlate with} older ages, as expected. \textnormal{This correlation is the result of star-formation quenching and natural aging processes (since progenitors of the most massive galaxies would have formed early according to hierarchical structure formation).} The slopes of the \age{} profiles are negative.
%, indicating that the quenching is being driven from the centers of the galaxies. 

\textnormal{Negative \age{} gradients are to be expected to some degree, as galaxies grow from the inside-out \citep[e.g.,][]{Sanchez21}, with stellar populations forming first near the center before star formation progresses to the outskirts.}
\textnormal{However, the gradients of the \age{} and \SFR{} profiles in Figures \ref{fig:BHmass}-\ref{fig:SFR_cumQM} become steeper as $M_{\rm BH}$ and \cumQM{} increase, except at the highest bin.}
%Steeper gradients for the \age{} profiles (listed in the figure legends) are indicative of stronger inside-out quenching. 
%For central galaxies, the gradients increase from low- to high-  $M_{\mathrm{BH}}$ and \cumQM{} until the second highest bin. 
Galaxies with the most massive black holes and highest imparted energies have older, flatter \age{} profiles\textnormal{ and are completely quenched in the \SFR{} profiles}. 
%This indicates that these galaxies have, on average, already been fully quenched. 
\textnormal{Since the brightest stars in a stellar population age and die quicker than dimmer populations, \age{} will increase faster in regions where star formation has been quenched more recently as compared to the \age{} of stellar populations that have been quenched for some time. This results in a flattening of the \age{} profiles once star formation in the outskirts quenches.}
\textnormal{Our results indicate that} the degree of inside-out quenching increases as $M_{\mathrm{BH}}$ and \cumQM{} increase, up to $M_{\mathrm{BH}}/M_{\odot} \approx 10^{8.5}$ and $\Sigma E_{\mathrm{QM}}\approx 10^{17}$, at which point the galaxy is completely quenched.%/[M_{\odot}\mathrm{kpc}^2 \mathrm{Gyr}^{-2}])\approx 17$.

This interpretation is supported by the distributions of main sequence, green valley, and quenched galaxies in these parameters, which are shown in Appendix \ref{app_corr}. Figures \ref{fig:cent_dist} and \ref{fig:sat_dist} show that very few main sequence galaxies have values of $M_{\mathrm{BH}}$ and \cumQM{} greater than the stated thresholds.

\textnormal{We note that there is some disagreement between our \SFR{} profiles in Figure \ref{fig:SFR_BHMass} and those obtained from MaNGA observations by \cite{Bluck2020b}. They found that central galaxies with $M_{BH}/M_{\odot}>10^{6.5}$ were, on average, quenched at small radii.  In contrast, we only see significantly quenched centers for TNG100 galaxies with central black hole masses $M_{BH}/M_{\odot}< 10^8$.  This disagreement of 1.5 orders of magnitude in SMBH mass for central quenching could be the result of a number of factors, including sample properties, comparability of $M_{BH}$ measurements in simulations and observations, or limitations with how black holes and feedback are modeled in the simulation. Despite quantitative disagreement, we find the same general trend as \cite{Bluck2020b} in the TNG100 high-mass central galaxies, with evidence for more inside-out quenching as black hole mass increases.}
%In Appendix \ref{app_SFR}, the \SFR{} profiles lie entirely below the quenching threshold for galaxies with $M_{\mathrm{BH}}/M_{\odot} > 10^{8.5}$ and $\Sigma E_{\mathrm{QM}}> 10^{17}$. There is some tension with the results of \cite{Bluck2020b}, who found completely quenched \SFR{} profiles for galaxies with $M_{\mathrm{BH}}/M_{\odot} > 10^{7.5}$, which we discuss more in Appendix \ref{app_SFR}. 

In the case of intrinsic parameters, it is challenging to disentangle the relative effects of mass and morphology on the distribution of star formation. The presence of a bulge correlates with the mass of a galaxy, which correlates with the star formation properties of a galaxy. Thus, the fact that galaxies in Figure \ref{fig:bulge}c,d) with \bulge$>0.5$ \textnormal{(yellow)} have older \age{} profiles may just be an indication that this population is, on average, more massive. There is generally more variability in the \age{} \textnormal{and \SFR{}} profiles when galaxies are binned by \bulge{} than by $M_{\mathrm{BH}}$ or \cumQM{}. This indicating that mass rather than morphology likely drives quenching in high-mass TNG100 galaxies.

Environmental quenching is not the dominant quenching process for high-mass galaxies, especially central galaxies. However, there are clear effects on the luminosity-weighted age in the outskirts of high-mass galaxies that reside in extreme environments. In Figure \ref{fig:halo}(c), central galaxies with $M_H/ M_\odot<10^{13.5}$ have flat \age{} profiles in the outskirts. At slightly higher halo masses, $10^{13.5}<M_H/M_\odot <10^{14}$ \textnormal{(red)}, the \age{} profile slope for central galaxies becomes positive in the outskirts ($R>1R_e$). The slope in the inner regions is still negative, indicating that this population is being quenched inside-out and outside-in. In the next highest halo mass bin, $10^{14}<M_H<10^{14.5}$ \textnormal{(brown)}, the profile for central galaxies is significantly flatter and older throughout than central galaxies with $10^{13.5}<M_H / M_\odot < 10^{14.5}$ \textnormal{(red)}.\textnormal{ In Figure~\ref{fig:SFR_halo}, the \SFR{} profiles of high-mass, central galaxy populations with $M_H/M_\odot> 10^{12.5}$ are completely quenched,} \textnormal{the same threshold identified by \cite{Bluck2020b} for all central galaxies}. \textnormal{For high-mass satellites, the halo mass threshold for quenching in Figure~\ref{fig:SFR_halo}d is $M_H/M_\odot>10^{13.5}$. This is slightly lower than the threshold identified by \cite{Bluck2020b} ($M_H/M_\odot>10^{14}$), although this threshold was determined with profiles constructed by satellite galaxies of all masses in their sample.}

Since halo mass is correlated with black hole mass \textnormal{for central galaxies}, it is instructive to compare \textnormal{\age{}} profiles at high $M_H$ with profiles at high $M_{BH}$. The typical galaxy with $M_{BH}/M_{\odot}>10^{8.5}$ is completely quenched, and the \age{} profile (Figure \ref{fig:BHmass}c) shows that this occurred through inside-out quenching. The population of galaxies with $10^{14}<M_H/M_\odot <10^{14.5}$ \textnormal{(brown)} is also completely quenched, but the inner gradient of the average \age{} profile (\grad $=-0.19\pm 0.04 \: {\rm dex/R_e}$) is much less negative than the same gradient for the $M_{BH}/M_\odot >10^{8.5}$ \textnormal{(Figure \ref{fig:BHmass}c, brown)} population (\grad $=-0.37 \pm 0.02 \: {\rm dex/R_e}$).

High halo masses (at $M_H/M_\odot \gtrsim 10^{13.5}$) are also correlated with high local galaxy overdensity. In the highest bin of $\delta_5$ that we include (Figure \ref{fig:overdens}c, \textnormal{red}), the slope of the \age{} profile for high-mass central galaxies becomes positive at $R\gtrsim 0.6 R_e$. \textnormal{This population is entirely quenched in profiles of \SFR{} (Figure~\ref{fig:SFR_overdens}).} The correlation between $M_H$ and $\delta_5$ makes it difficult to ascertain which mechanisms $-$ interactions with the hot gas halo or interactions with other galaxies $-$ are acting to quench the outskirts of these systems. %\textnormal{The \SFR{} profiles of high-mass central galaxies are completely quenched at $\delta_5>1$}

Similar trends are present, but less pronounced, for high-mass satellite galaxies that reside in similar environments (Figures \ref{fig:halo}d and \ref{fig:overdens}d). This may be due to the fact that some satellites in the most massive halos, or with the largest values of $\delta_5$, may have joined the system only recently. For high-mass satellites with $M_H/M_\odot>10^{13.5}$, the \age{} profiles are similar at $R\lesssim 0.5R_e$, but the outskirts become older with increasing halo mass. The slopes of \textnormal{the \age{}} profiles of high-mass satellites with $\delta_5>10^{1.5}$ become slightly positive at $R\gtrsim 1.1R_e$. \textnormal{The \SFR{} profiles of high-mass satellites subdivided by $\delta_5$ are very noisy (Figure \ref{fig:SFR_overdens}d), but appear mostly quenched at $\delta_5>10^{1.5}$. In \cite{Bluck2020b}, \SFR{} profiles of high-mass satellite galaxies are completely quenched at $\delta_5>10^{1}$. Additionally, we find that high-mass TNG100 satellites in the most underdense regions ($\delta_5<10^0$, purple) are completely quenched, while MaNGA satellites in similar regions maintain star formation throughout. We note that our threshold for high-mass galaxies is higher than that of \cite{Bluck2020b}, which defined high-mass satellites to have $M_*/M_\odot> 10^{10}$. We further discuss the tension between our results and that of \cite{Bluck2020b} for satellites subdivided by $\delta_5$ in the following subsection.}

The radial profiles of \age{} for high-mass galaxies, subdivided by $M_H$ and $\delta_5$, indicate that environmental quenching can operate on these systems, suppressing star formation from the outside-in. \textnormal{At the masses and densities where environmental quenching affects the \age{} profiles, galaxies are already completely quenched in the \SFR{} profiles.} \textnormal{While extreme environments can contribute to the quenching of high-mass galaxies,} the dominant quenching pathway for high-mass galaxies is via AGN feedback energy, as we conclude above. 

%An analogous reduction in \SFR{} is not seen in Appendix \ref{app_SFR}, as these galaxies are already completely quenched. Luminosity-weighted age traces out star formation on longer timescales than \SFR{}, indicating that these environmental processes operated on central galaxies in the past. %ADD?: Peng 2012 found that environmental quenching only occurred in satellite galaxies, can I clarify those results?

%High-mass satellites that joined at early times are generally flatter and older than those that joined later. However, the population of high-mass satellites with $1.<z_j<1.5$ is an exception, with a profile that is younger than the population of high-mass satellites with $0.75.<z_j<1.$. It is unclear why this should be the case, and this phenomenon might disappear with a larger sample, as only $49$ high-mass satellites contribute to the $1.<z_j<1.5$ population.

%Environmental effects also explain why satellites split by $M_{\mathrm{BH}}$ and \cumQM{} in Figures \ref{fig:BHmass} and \ref{fig:cumQM} generally have similar ages toward the center but older ages and flatter profiles in the outskirts, when compared to central galaxies with identical $M_{\mathrm{BH}}$ and \cumQM. However, mass quenching is still the dominant process in high-mass satellite galaxies.

In Appendix \ref{app_corr}, we provide further evidence that environmental processes are not the dominant quenching process for high-mass satellite galaxies. However, environment can still impact the evolution of these systems. High-mass satellites that joined their central galaxies at earlier times have \age{} profiles that are generally older and flatter than those for satellites that at joined later times, although the population with $1.0<z_j<1.5$ \textnormal{(yellow)} is a curious exception \textnormal{in both \age{} and \SFR{} profiles}. High-mass satellites that have lost any amount of star-forming gas since joining are, on average, completely quenched. The \age{} profiles of these systems have negative gradients, indicating that the quenching was driven by the central SMBHs. Profiles of \age{} for the population of high-mass satellite galaxies that have gained star-forming gas also show evidence of inside-out quenching \textnormal{(Figure~\ref{fig:SFgas}b, green)}, but the outskirts are significantly younger than populations with negative \sfgas. It is clear, especially when considering the corresponding profiles of \SFR{} in \textnormal{Figure~\ref{fig:SFR_SFgas}}, that the enhancement of star-forming gas has operated from the outside-in.

To summarize this section, the radial profiles of high-mass galaxies are primarily shaped by mass-quenching operating inside-out. However, environmental processes can affect the shape of \age{} profiles and contribute to quenching at large radii in extreme environments. 

\subsection{Low-mass galaxies}

 The typical \age{} profile for low-mass TNG100 galaxies appears similar to the \age{} profile for disk-dominated galaxies with $10^{9.2}< M_*/M_{\odot} <10^{10.2}$ observed by CALIFA by \cite{Gonzalez14},\textnormal{ including the dip in \age{} at the center. From visual inspection of maps of $\Sigma_{\rm SFR}$ in low-mass galaxies, active star formation in the center of these galaxies is common. This is generally not reflected in profiles of \SFR{} and sSFR in this work and observational analysis, likely due to the relatively high stellar mass density at the centers of galaxies. Further exploration of star formation in the centers of low-mass galaxies is desirable, especially since AGN feedback in the TNG simulations is imparted at the centers of galaxies. However, such explorations are limited by resolution in both simulations and observations.} 
 
 Both simulated and observed low-mass galaxies have shallower profiles than corresponding high-mass galaxies, indicating less influence by the SMBH. \textnormal{The \age{} and \SFR{} gradients that are present reflect the natural growth of the galaxies and the aging of their stellar populations.} There is no significant difference in \age{} or \SFR{} profile shape or normalization between simulated low-mass galaxy populations over two magnitudes of black hole mass ($10^{6} < M_{BH}/M_{\odot} <10^8$). 

There is no \age{} or \SFR{} profile dependence on \cumQM{} for simulated low-mass central galaxies, with a slight exception for galaxies with $10^{13}<\Sigma E_{QM} <10^{15}$ at small radii. For low-mass satellite galaxies, however, there is a clear difference in normalization. In \textnormal{Figure~\ref{fig:SFR_cumQM},} satellite galaxies with $10^{13}< \Sigma E_{QM} <10^{15}$ (green) are entirely quenched except at the smallest radii. Central galaxies with comparable \cumQM{} are still forming stars at the same \SFR{} as central galaxies with negligible \cumQM{}, albeit with greater variation. There is no compelling reason to expect that the same amount of AGN energy imparted into a satellite galaxy would quench the satellite but not a central of similar mass. It may be that this is a matter of correlation, rather than causation. %Additionally, here we have only investigated a sample of low-mass satellite galaxies that have a relatively high \cumQM.% where the SMBH is being fed by FLESH OUT AFTER LOOKING AT FIGURE (env. effects backed up by flatter outskirt profile)

It is unclear what role a bulge-like morphology plays in quenching or regulating star formation in low-mass TNG100 galaxies. As discussed above, the strong correlation of \bulge{} with other galaxy parameters complicates this investigation. There is certainly a dependence on \bulge{} for \age{} profiles of TNG100 galaxies, but this is not fully reflected in the \SFR{} profiles in \textnormal{Figure ~\ref{fig:SFR_bulge}}. \textnormal{The flatter shapes of the \age{} profiles of low-mass galaxies with \bulge$>0$ are likely driven by the aging of stellar populations in the bulge, especially since there are not significant differences in the \SFR{} profiles of low-mass galaxies separated by \bulge.} %The \age{} profile differences are likely due to the stellar populations present in the bulge, which dominates the center out to $\sim0.5 R_e$ CITE KALINOVA. 
There is evidence from observations that environmental effects, particularly ram pressure stripping, can drive morphological transformation as well as the availability of star-forming gas in satellite galaxies \citep[e.g.,][]{Steyrleithner20,Marasco23}. Thus, morphology may not drive star formation properties, but could be affected by the same evolutionary processes.

As quenching in low-mass TNG100 galaxies is not being driven by intrinsic factors, we therefore turn to environmental processes to explain the quenched low-mass galaxies in our sample. Here, we categorize galaxies into star-forming type in the same way as in \cite{McDonough23}. Quenched galaxies are those that are $\ge 1.1$~dex below the star-formation main sequence on a $M_*-\mathrm{SFR}$ diagram. We can expect differences in central and satellite galaxy quenching due to the different environmental effects these systems will experience. The likelihood of galaxy-galaxy interactions for both central and satellite galaxies will increase with increasing local galaxy density.  However, only satellite galaxies will experience significant galaxy-halo interactions along their orbits within the gas halos.

\begin{figure}
    \centering
    \includegraphics[scale=0.8]{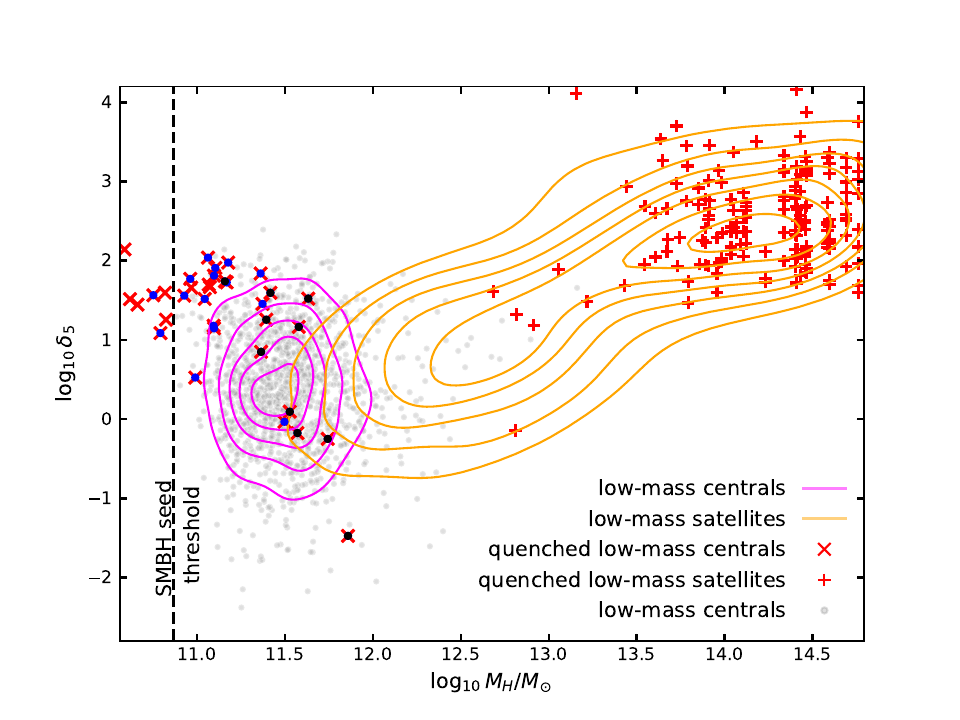}
    \caption{Comparison of the distribution of quenched low-mass central and satellite galaxies to the entire population of low-mass central and satellite galaxies. Magenta and orange contours indicate the density of the entire low-mass central and satellite population, respectively. All individual low-mass central galaxies are plotted as grey points. Quenched low-mass central galaxies are indicated as red crosses and quenched low-mass satellite galaxies are indicated as red plus signs. \textcolor{black}{Low-mass, quenched centrals are overplotted with a blue dot if they contain any SMBH and a black dot if they contain an SMBH with ten times the SMBH seed mass. The black dashed line indicates the halo mass threshold for SMBH seeding.}}
    \label{fig:halo-dens}
  
\end{figure}

The \age{} profiles of low-mass central galaxies, subdivided by $M_{H}$ and $\delta_5$, deviate only slightly from each other at high halo masses and local galaxy overdensities (Figures \ref{fig:halo}a and \ref{fig:overdens}a). The only exception is for central galaxies with $M_H/M_\odot <10^{11}$ \textnormal{(purple)}, which are significantly older than central galaxies in higher mass halos. 
In total, there are $40$ low-mass, quenched, central galaxies in our sample.
%, only $13$ of which have non-negligible \cumQM. 
The median values of $M_{H}$ and $\delta_5$ for the population of low-mass, quenched centrals ($\log_{10} M_{H}/M_{\odot}=11.14$, $\log_{10} \delta_5 = 1.52$) differ from those of all low-mass centrals ($\log_{10} M_{H}/M_{\odot}=11.49$, $\log_{10} \delta_5 = 0.405$).
%In other words, quenching in low-mass central galaxies tends to occur at higher local galaxy overdensities and lower halo masses. 

In Figure \ref{fig:halo-dens}, we draw contours around the density of low-mass central (magenta) and satellite (orange) galaxies in the $M_{H}-\delta_5$ parameter space. Quenched central (crosses) and satellite (plus signs) galaxies are individually plotted in red. The halo mass threshold for SMBH seeding in TNG100 is shown as a black dashed line. A blue point is overplotted on points representing quenched, low-mass centrals that contain any SMBH, and black points are plotted for quenched, low-mass centrals with SMBHs that are more than ten times the SMBH seed mass. The majority of quenched low-mass centrals lie outside the density contours for the population of all low-mass centrals and are biased toward low halo masses. However, their quiescence in TNG100 is likely due to lacking an SMBH, at least until recent times. Without an SMBH to regulate star formation through AGN feedback, these systems over-efficiently convert gas to stars throughout their early lives and are quiescent at late times. 
%However, there are other low-mass central galaxies with similar values of $M_H$ and $\delta_5$ and are not quenched. This is why the profile for low-mass central galaxies with $\delta_5 >10^{1.5}$ in Figure \ref{fig:overdens} differs only slightly in normalization, since much of the population remains star-forming. 

There are a small number of low-mass, quenched centrals with more typical halo masses that have an SMBH that is more than ten times the TNG100 SMBH seed mass, meaning that the SMBHs have been accreting for some time. With such a small sample and without further analysis of these systems, it is impossible to determine how these systems became quiescent. One possibility is galaxy-galaxy interactions that can either efficiently strip gas from a system or trigger a starburst phase, during which a galaxy consumes most of its available gas over a short period of time, leading to a post-starburst quiesence \citep[e.g.,][]{Wilkinson21}. The likelihood of such interactions is expected to scale with local galaxy overdensity, and thus the fraction of low-mass centrals that are quenched should increase with increasing $\delta_5$. However, our sample is too small to perform such an analysis.
%It is not $\delta_5$ that directly affects the availability of star-forming gas, but rather the interactions that scale with $\delta_5$. Thus, it may be that quenching in low-mass central galaxies is driven not by the number of interactions, but rather the nature of interactions that take place. It is possible that the quenched galaxies in our sample underwent interactions that ultimately resulted in quiescence (e.g., interactions that result in a starbursting phase that drives stellar feedback and uses up available gas, or interactions that otherwise efficiently removed gas available for star-formation). The lower gravitational potential for galaxies in low-mass halos would make it easier for gas to be stripped from these system. As $\delta_5$ only indicates the odds of interactions occurring, the few low-mass, quenched, central galaxies with more typical values of $\delta_5$ and $M_H$ could still have been quenched by interactions.

Unlike central galaxies, there is no correlation between $M_{H}$ and stellar mass for satellite galaxies  (i.e., here $M_{H}$ is the mass of the halo within which the satellite is orbiting). However, interactions with the hosts' hot gas halo and host halo potential, such as ram pressure stripping and dynamical friction, should scale with halo mass \citep[e.g.,][]{Woo15,Bluck16}. The \age{} profiles of simulated, low-mass satellite galaxies in the most massive halos ($M_{H}>10^{13.5} M_\odot$) and the most overdense regions ($\delta_5>10^{1.5}$) have higher normalizations than the \age{} profiles of low-mass satellites in less massive halos and less overdense regions. This is especially prominent in the profiles of populations subdivided by halo mass (Figure \ref{fig:halo}b), where satellites in halos with $M_H/M_\odot >10^{14}$ are $\sim 0.5$~dex older at $R \gtrsim 0.25 R_e$ than satellites in $M_H/M_\odot <10^{13.5}$ halos. \textnormal{In Figure~\ref{fig:SFR_halo}b), the only low-mass satellite galaxies that are quenched at any radii are those with $M_H/M_\odot >10^{14}$. This threshold is the same as that identified by \cite{Bluck2020b} for MaNGA satellites, although they found that low-mass satellites at halo masses greater than this threshold had \SFR{} profiles that were entirely quenched. \cite{Bluck2020b} also found that the \SFR{} profiles of satellites are systematically higher than those for central galaxies at all halo masses. We do not find this to be the case for TNG100 satellites once the satellites are subdivided by stellar mass. }

It has been shown previously in observations that the quenched fraction of satellite galaxies increases with local galaxy overdensity and halo mass \citep[e.g.,][]{Bluck16,Bluck2020a}. In Figure \ref{fig:halo-dens}, we show that quenched, low-mass satellites generally reside well within the distribution of all low-mass satellites, although they are clustered at the high $M_{H}$ end of the distribution. \textnormal{When subdivided by $\delta_5$, we find that there are no populations of low-mass satellite galaxies where the average \SFR{} profiles have any quenched regions. This is in tension with \cite{Bluck2020b}, which found that low-mass satellites at $\delta_5>10^{1.5}$ have \SFR{} profiles that lie completely below the quenching threshold, but there are several caveats to this comparison. There are resolution limitations in how galaxy-galaxy interactions are modeled in the simulation \citep[see, e.g.,][]{Sparre16}. Additionally, there are also differences in how local galaxy overdensity is measured: in this work, $\delta_5$ is measured with 3D positions, while observations are limited by the 2D view of the sky.}

The longer a satellite spends in a halo, the more time there is for environmental interactions to take place, which motivates our investigation of the dependence of radial profiles on $z_j$. In Figure \ref{fig:joinz} (top), low-mass satellites that joined their hosts' halos earlier have, on average, older \age{} than systems that joined their hosts' halos only recently. In Figure \ref{fig:SFR_joinz}, we see that only the populations of low-mass satellite galaxies that joined earlier than $z=1$ are, on average, almost completely quenched. Additionally, beyond $R\approx 0.5 R_e$, the \age{} profiles of these early joiners are flatter than they are for low-mass satellites that joined later. While this points to outside-in quenching operating in these systems, the \SFR{} profile of the population with $1.0<z_j<1.5$ \textnormal{(Figure~\ref{fig:SFR_joinz})} shows evidence of active star formation occurring at $\approx 1.25R_e$, although this population is highly variable. 

The redshift at which a satellite joined its host serves as a statistical proxy for cumulative environmental interactions the satellite has experienced. In contrast, \sfgas{} is a direct measurement of the net effect of these interactions on the total mass of available star forming gas. It is no surprise, then, that \age{} profiles subdivided by \sfgas$/M_*$ are clearly separated in age. Low-mass satellites with \sfgas$/M_*>0.25$ have a total median luminosity-weighted age of $\approx 370$ Myr, while low-mass satellites with \sfgas$/M_*<-0.25$ have a total median \age{} of $\approx 6.7$ Gyr \textnormal{and are completely quenched in the outskirts (Figure~\ref{fig:SFR_SFgas}a))}. 

\begin{figure}[ht]
    \centering
    \includegraphics[scale=0.8]{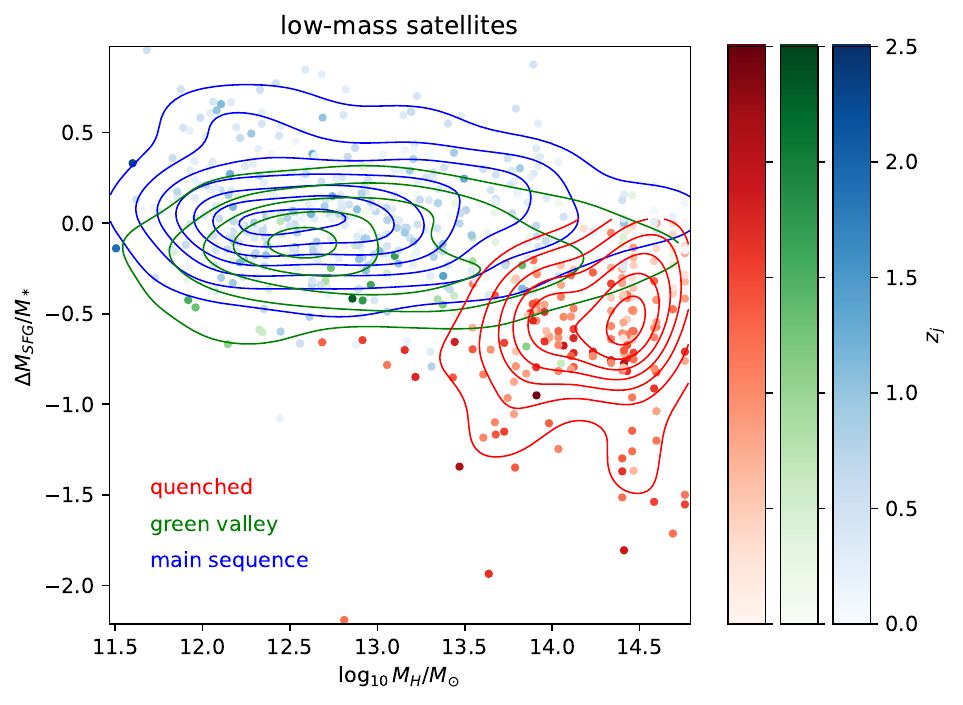}
    \caption{The relationship between halo mass and change in mass of star forming gas since joining, for main sequence (blue), green valley (green), and quenched (red) low-mass satellites. Density contours are drawn for the separate populations, and individual galaxies are plotted as points that are colored according to their star-formation type and joining redshift, $z_j$.}
    \label{fig:halo-sfgas}
  
\end{figure}

What drives the change in star-forming gas mass for low-mass TNG100 satellite galaxies? Of all the parameters explored in Appendix \ref{app_corr}, \sfgas$/M_*$ is best correlated with $z_j$, although it is not a tight correlation. However, the redshift at which a satellite joined its host's halo is not indicative of the actual physical processes that are occurring. In Appendix \ref{app_corr}, we show correlations with \sfgas$/M_*$ for $z_j$, $M_{H}$, and $\delta_5$, with satellites separated by mass and star formation type (Figure~\ref{fig:orbitcorr}). In Figure \ref{fig:halo-sfgas}, we show the most significant correlation for low-mass satellites: \sfgas$/M_*$ and $M_{H}$. Most quenched, low-mass satellites joined high-mass halos relatively early and have lost significant amounts of gas by the present day. The quenched population has greater overlap with main-sequence and green valley galaxies in the \sfgas$/M_* - \delta_5$ parameter space (see Figure \ref{fig:orbitcorr} in Appendix \ref{app_corr}). Thus, we can infer that quenching of low-mass TNG100 satellite galaxies is driven more typically by environmental processes related to their hosts' halos, rather than galaxy-galaxy interactions that scale with $\delta_5$. Host halos have a greater chance of removing gas from the satellites over longer timescales, driving the correlation of quenching with early $z_j$. In a study of SDSS-MaNGA satellites, \cite{Oyarzun23} reached similar conclusions regarding the environmental quenching of observed low-mass satellites in high-mass halos over time.

It is important to note that this is a very general picture, and there are many exceptions, as can be seen in Figure \ref{fig:halo-sfgas}. In some cases, TNG100 satellites that joined their hosts only recently can still lose large amounts of star-forming gas. In addition, there are satellites that joined their hosts in the distant past, but remain actively star forming at the present day, with some having even gained star-forming gas. We conclude that gas removal and quenching in low-mass satellites tends to be dominated by interactions with their hosts' halos and is dependent on the exact nature of the interactions that individual satellites experience. This interpretation is consistent with results from the FOGGIE simulations, which have shown that ram pressure stripping of satellites at $z=2$ is stochastic and highly dependent on the paths of individual satellites through the circumgalactic medium \citep{Simons20}. Additionally, using the cosmological hydrodynamic code Enzo \citep{Tonnesen07} to simulate a massive cluster, \cite{Tonnesen08} found that the ram pressure force exerted on satellites at the same cluster-centric distance could vary by over an order of magnitude due to differences in local density and relative speed. \textnormal{Finding relatively high metallicities in rapidly quenched, low-mass satellite galaxies, the results of \cite{Corcho-Cabellero2023} indicate that these systems are quenched following a starbursting period triggered by environmental interactions.} Regardless of how it occurs, radial profiles of \age{} and \SFR{} show that environmental quenching occurs from the outside-in in TNG100 galaxies, in agreement with observations, e.g., \cite{Finn18}, \cite{Lin19}, \cite{Bluck2020b}, and \cite{Matharu21}. 

Our conclusion that low-mass satellite quenching is more dependent on halo mass than it is on local galaxy density is in tension with the results of \cite{Bluck2020b}. Using a random forest classification analysis, \cite{Bluck2020b} found that local galaxy overdensity was most predictive of quenched spaxels in low-mass satellite galaxies observed by MaNGA. Halo mass was the second most important environmental parameter, and distance to the host galaxy was found to be of little importance. We note, however, that measurements of $M_H$ and $\delta_5$ are much more challenging in observations than they are in simulations. In the case of \cite{Bluck2020b}, the halo masses were derived from abundance matching techniques, which could result in a stronger correlation of $M_H$ with $\delta_5$ than is physical. However, this tension could also result from limitations in the simulation, for example the spatial resolution at which the gas physics is captured, particularly during mergers or other interactions \citep{Sparre16}. \textnormal{Further work in both observations and simulations would be necessary in order to determine whether the primary driver of low-mass satellite quenching is galaxy-galaxy interactions (correlated with $\delta_5$) or galaxy-halo interactions (correlated with $M_H$). We remind readers that simulations are based on theoretical models, and may not represent reality.}
%derive halo masses for simulated galaxies in an observationally-comparable manner
%repeat random forest analysis done by Bluck

%Our results have a number of implications for sample selection in future studies that examine environmental quenching. Firstly, quenching in high-mass galaxies is driven by intrinsic processes that must be accounted for. 
%Low-mass galaxies are better suited for studies of environmental quenching, since quenching in these systems is driven by environmental processes, which is not the case for high-mass galaxies.
%Our results imply that, in the TNG100 simulation,  the quenching of low-mass satellite galaxies tends to be driven by interactions with the host gas halo.  
%\textcolor{red}{For most low-mass central galaxies, the driving force behind quiescence in the simulation appears to be a lack of SMBH due to the TNG100 seeding procedure. Further work would be necessary to determine the cause of quiescence in the few low-mass, quenched centrals that have hosted a SMBH for some time, but galaxy-galaxy interactions may play a role. }
%Low-mass central and satellite galaxies are typically quenched by different environmental processes.
%, and quenching may depend more on the nature of these stochastic processes than \textcolor{violet}{the rate of interactions} as estimated by local galaxy overdensity. 
%The efficiency with which environmental processes (e.g., ram pressure stripping and galaxy-galaxy interactions) remove the gas reservoir appears to be highly dependent on the nature of individual interactions, which may bias results in small samples.

\section{Summary} \label{sec:summary}

In this paper, we constructed radial profiles of \age{} \textnormal{and \SFR{}} for various populations of galaxies in the TNG100 simulation in order to investigate the degree to which the profiles are affected by intrinsic and environmental factors. We constructed these profiles separately for central and satellite galaxies with low- and high- masses (\lomass{} and \himass{}, respectively). Our main results for galaxies in the TNG100 simulation are as follows:

\begin{itemize}
    \item The shapes and normalizations of averaged radial profiles of high-mass (\himass) galaxies (whether central or satellite) are strongly dependent on black hole mass ($M_{BH}$, Figures \ref{fig:BHmass} \textnormal{and \ref{fig:SFR_BHMass}}) and cumulative AGN feedback imparted on the galaxies (\cumQM, Figures \ref{fig:cumQM} \textnormal{and \ref{fig:SFR_cumQM}}). The shapes of these profiles are consistent with inside-out quenching.

    \item High-mass galaxies in extreme environments (Figure \ref{fig:halo}: $ M_H/ M_{\odot}>10^{13.5}$ for central galaxies and $ M_H/ M_{\odot}>10^{14}$ for satellite galaxies and/or Figure \ref{fig:overdens}: $ \delta_5 >10^{1.5}$) have \age{} profiles that become slightly positive at large radii. This indicates that galaxies in extreme environments can be quenched from both inside-out and outside-in.

    \item Population-averaged \age{} \textnormal{and \SFR{}} profiles of low-mass galaxies are similar across several magnitudes of black hole mass. This is also the case for low-mass central galaxies when binned by \cumQM{} (Figures \ref{fig:cumQM} \textnormal{and \ref{fig:SFR_cumQM}}). When low-mass satellites are binned by \cumQM, \age{} normalization increases with increasing \cumQM{} \textnormal{and \SFR{} decreases}. The averaged \age{} profile for low-mass satellites with $10^{13}<$\cumQM$<10^{15}$ is flatter in the outskirts compared to the other low-mass satellite \cumQM{} populations, \textnormal{and the \SFR{} profile of this population is consistent with outside-in quenching}. 

    \item The \age{} \textnormal{and \SFR{}} profiles of low-mass central galaxies are not strongly dependent on $\delta_5$ or $M_H$ (Figures \ref{fig:halo}-\ref{fig:SFR_overdens}). The only exceptions are central galaxies in very low-mass halos ($M_H/M_\odot <10^{11}$), which are much older than low-mass centrals in higher mass halos. 
    This is likely the result of overcooling because the TNG100 model only seeds a SMBH in a galaxy once the halo mass reaches a threshold of $M_H/M_\odot \approx 10^{10.87}$. Only low-mass satellites in extreme environments ($M_H/ M_{\odot}>10^{13.5}$ and/or $\delta_5>10^{1.5}$) have \age{} \textnormal{and \SFR{}} profiles that differ significantly from the other \textnormal{low-mass satellite} populations. 

    \item The time a satellite has spent within its host's halo, $z_j$, has a greater effect on \age{} \textnormal{and \SFR{}} profiles of low-mass satellites than it does for high-mass satellites (Figures \ref{fig:joinz} \textnormal{and \ref{fig:SFR_joinz}}). The normalization of \age{} for low-mass satellites increases with increasing time in the halo \textnormal{and quenched regions appear in the \SFR{} profiles at $z_j>1.0$. }

    \item While $z_j$ serves as an imperfect proxy for environmental effects a satellite encounters in a host's halo, \sfgas{} directly measures the change in star-forming gas that a satellite has experienced since joining. The change in star-forming gas preferentially affects the outskirts of both high- and low-mass galaxies. For low-mass satellite galaxies, the shapes and normalizations of \age{} \textnormal{and \SFR{}} profiles are highly dependent on \sfgas.
    
    %\item  \textcolor{red}{Quenching in low-mass galaxies is not driven by AGN feedback, but by stochastic environmental processes. We find that many quenched low-mass centrals reside in unusually low-mass halos and in regions of space that have high galaxy overdensities (Figure \ref{fig:halo-dens}). We infer that galaxies in low-mass halos were more susceptible to being stripped of gas and were quenched via galaxy-galaxy interactions. }

    \item Quenched low-mass satellites are found almost entirely within high-mass halos ($M_H/M_{\odot}>10^{13.5}$, Figure \ref{fig:halo-dens}). Quenched low-mass satellites are most distinct as a population in $M_H$-\sfgas{} parameter space (Figure \ref{fig:halo-sfgas}). We infer that quenching in low-mass \textnormal{TNG100} satellites occurs when interactions with massive host halos deplete these galaxies of star-forming gas. These interactions are more likely to occur, and remove more gas, with increasing time spent in the halo. However, $z_j$ and \sfgas{} are only loosely correlated with each other (Figure \ref{fig:sat_dist}). Thus, the efficiency with which environmental processes (e.g., ram pressure stripping and galaxy-galaxy interactions) remove the gas reservoir appears to be highly dependent on the nature of individual interactions.
    
\end{itemize}

In conclusion, we find that high-mass galaxies, both centrals and satellites, in the TNG100 simulation are quenched from the inside-out via AGN feedback energy, in agreement with expectations based on previous observational and theoretical studies of quenching. %Quenching in low-mass centrals is uncommon, and is likely driven by interactions with other galaxies. On the other hand, 
Quenching in low-mass TNG100 satellites appears to be driven by interactions with their hosts' hot gas halos. These low-mass quenched populations are not always reflected in the average radial profiles of different environmental populations; the parameters we have explored only capture the likelihood of interactions occurring, rather than the natures of the interactions. Environmental effects drive quenching from the outside-in, resulting in radial \age{} profiles that are positive in slope beyond $R\sim 1-1.25 R_e$ \textnormal{and \SFR{} profiles that are quenched at large radii}. We have shown previously that the radial profiles of \age{} \textnormal{and \SFR{}} for TNG100 galaxies are generally in agreement with observations \citep{McDonough23}. However, the results we report here should still be interpreted with the understanding that simulations are limited by resolution, size, and the physical models that are employed.

\begin{acknowledgements}
    \textnormal{We thank the reviewer for providing detailed feedback on our work.} We would like to acknowledge the work and documentation provided by the IllustrisTNG team that has made this paper possible. The IllustrisTNG simulations were undertaken with compute time awarded by the Gauss Centre for Supercomputing (GCS) under GCS Large-Scale Projects GCS-ILLU and GCS-DWAR on the GCS share of the supercomputer Hazel Hen at the High Performance Computing Center Stuttgart (HLRS), as well as on the machines of the Max Planck Computing and Data Facility (MPCDF) in Garching, Germany. The additional computational work done for this paper was performed on the Shared Computing Cluster which is administered by Boston University’s Research Computing Services. This work was partially supported by NSF grant AST-2009397. \textnormal{BM acknowledges support by Northeastern University's Future Faculty Postdoctoral Fellowship Program. We also thank Dr. Paul Tol for providing an online resource for sequential color maps that are colorblind-friendly.}
\end{acknowledgements}

%% To help institutions obtain information on the effectiveness of their 
%% telescopes the AAS Journals has created a group of keywords for telescope 
%% facilities.
%
%% Following the acknowledgments section, use the following syntax and the
%% \facility{} or \facilities{} macros to list the keywords of facilities used 
%% in the research for the paper.  Each keyword is check against the master 
%% list during copy editing.  Individual instruments can be provided in 
%% parentheses, after the keyword, but they are not verified.

\vspace{5mm}
%\facilities{HST(STIS), Swift(XRT and UVOT), AAVSO, CTIO:1.3m,
%CTIO:1.5m,CXO}

%% Similar to \facility{}, there is the optional \software command to allow 
%% authors a place to specify which programs were used during the creation of 
%% the manuscript. Authors should list each code and include either a
%% citation or url to the code inside ()s when available.

\software{astropy \citep{astropy:2013,astropy:2018, astropy:2022}
          }

%% Appendix material should be preceded with a single \appendix command.
%% There should be a \section command for each appendix. Mark appendix
%% subsections with the same markup you use in the main body of the paper.

%% Each Appendix (indicated with \section) will be lettered A, B, C, etc.
%% The equation counter will reset when it encounters the \appendix
%% command and will number appendix equations (A1), (A2), etc. The
%% Figure and Table counter will not reset.

%% For this sample we use BibTeX plus aasjournals.bst to generate the
%% the bibliography. The sample631.bib file was populated from ADS. To
%% get the citations to show in the compiled file do the following:
%%
%% pdflatex sample631.tex
%% bibtext sample631
%% pdflatex sample631.tex
%% pdflatex sample631.tex

\bibliography{bib}{}
\bibliographystyle{aasjournal}

\appendix

\section{Inter-correlation of parameters} \label{app_corr}

\begin{figure}[hb!]

    \centering
    \includegraphics[height=\textwidth, keepaspectratio]{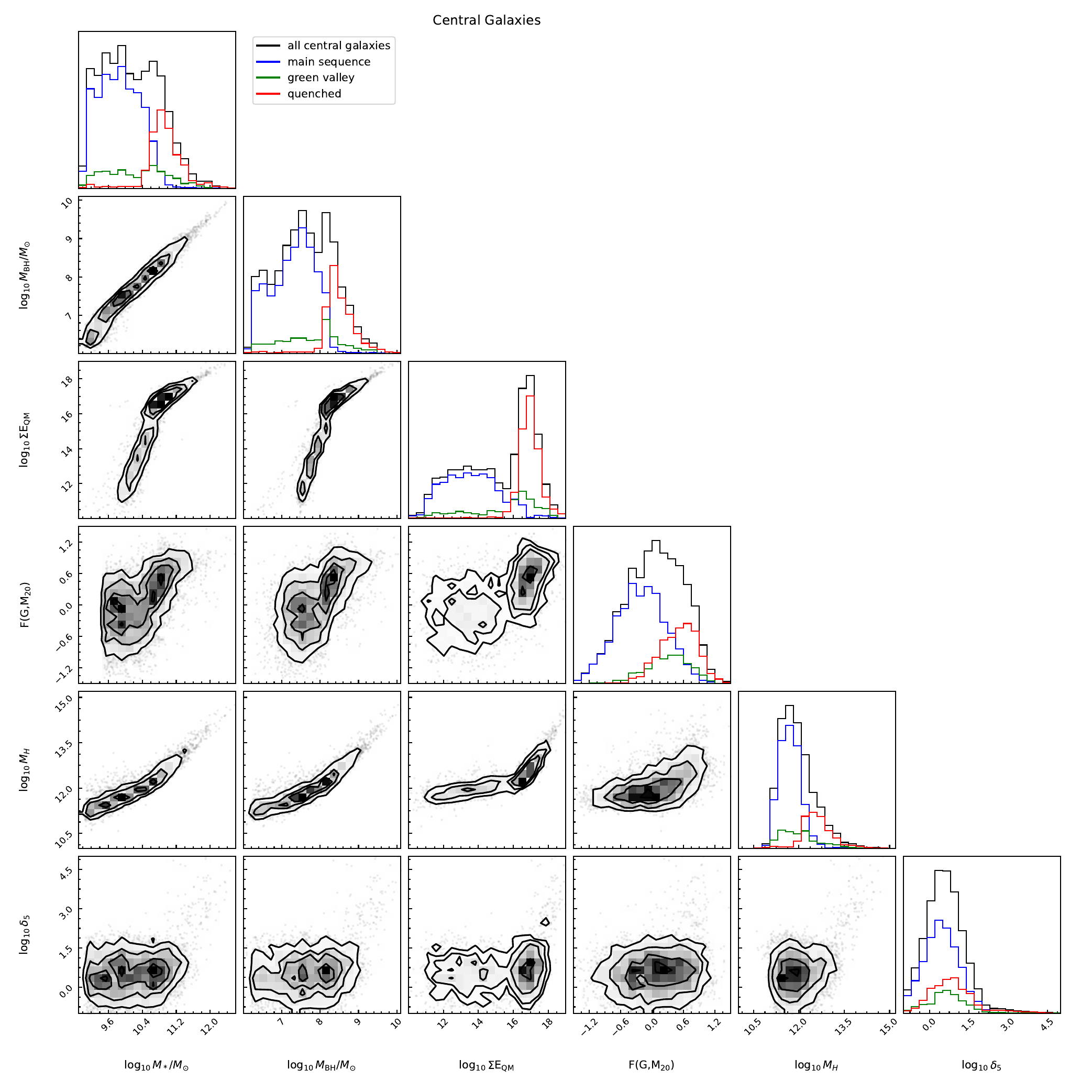}
    \caption{Corner plot illustrating correlations among the parameters explored in this paper, for central galaxies only. The diagonal displays the total distribution of each parameter (black), and the distribution separated by galaxy star-formation type: main sequence (blue), green valley (green), and quenched (red). Off-diagonal plots are two-dimensional histograms revealing relationships between two parameters. Contours are drawn at (0.5, 1, 1.5, 2)-sigma levels. }
    \label{fig:cent_dist}
  
\end{figure}

\begin{figure}[ht!]
    \centering
    \includegraphics[height=\textwidth, keepaspectratio]{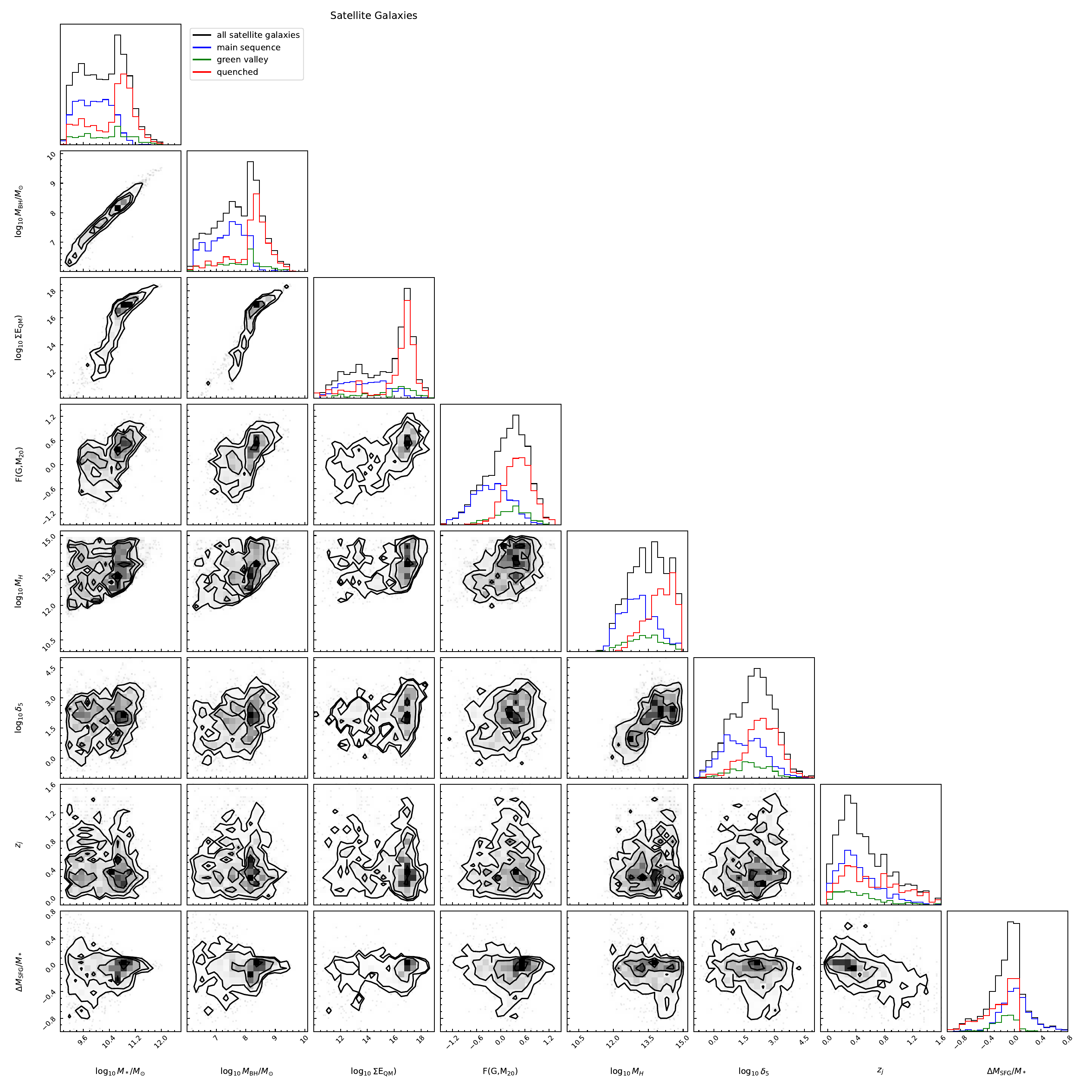}
    \caption{Corner plot illustrating correlations among the parameters explored in this paper, for satellite galaxies only. Formatting, including parameter ranges, is identical to Figure \ref{fig:cent_dist}, with the addition of two parameters not computed for central galaxies: z$_j$ and \sfgas.}
    \label{fig:sat_dist}
  
\end{figure}

Galaxy evolution is a complex process that is by no means fully understood at this time. Adding to the inherent complexity of the galaxy evolution process is that parameters that describe properties of galaxies are often correlated, either through causation or co-evolution. Because of this, it is important to understand how the parameters we have explored in this paper are related to one another. In Figures \ref{fig:cent_dist} and \ref{fig:sat_dist}, we present corner plots showing the relationships between the parameters we explored for central and satellite galaxies, respectively. On the diagonals, the one-dimensional distribution of each parameter is shown for all galaxies (black), and separately for galaxies based on their star-formation activity. Adopting the definitions from \cite{McDonough23}, we show the distributions for galaxies that are classified as quenched (red), green valley (green), or on the main sequence (blue).

Figure \ref{fig:cent_dist} shows the tight linear correlation between galaxy stellar mass and black hole mass, $M_* - M_{BH}$, for TNG100 central galaxies. Stellar mass is also correlated with total cumulative AGN energy imparted in the quasar mode, total halo mass, and, to a lesser extent, the bulge statistic. The existence of these correlations is well-documented in the literature \citep[e.g.,][]{Beifiori12,Reines15,Wechsler18}. On average, more massive TNG100 galaxies have more massive black holes, more massive halos, and more bulge-like morphologies. The total energy imparted by a SMBH is not directly observable, but it is expected to correlate well with black hole mass, as the imparted energy is related to the accretion rate. More massive galaxies are also more likely to be quenched. 

On the diagonal in Figure \ref{fig:cent_dist}, the distributions of quenched and main sequence central galaxies are distinct for all parameters related to stellar mass. At higher stellar masses, more galaxies are quenched. This phenomenon is thought to be driven by the central SMBH, and is often referred to as `mass quenching' \citep{Peng10}. The distribution is especially bimodal for cumulative energy injected, where quenched central galaxies lie almost entirely in the range of \cumQM=$10^{16}-10^{19}$. This provides further evidence that mass quenching is driven by the energy imparted by the SMBH. 

Local galaxy overdensity is not correlated with stellar mass or other parameters, except at the highest galaxy masses. The most massive central galaxies will only exist in dense cluster centers, where they can accumulate stellar mass through accretion.

\begin{figure}[htb!]
    \centering
    \includegraphics[width=\textwidth, keepaspectratio]{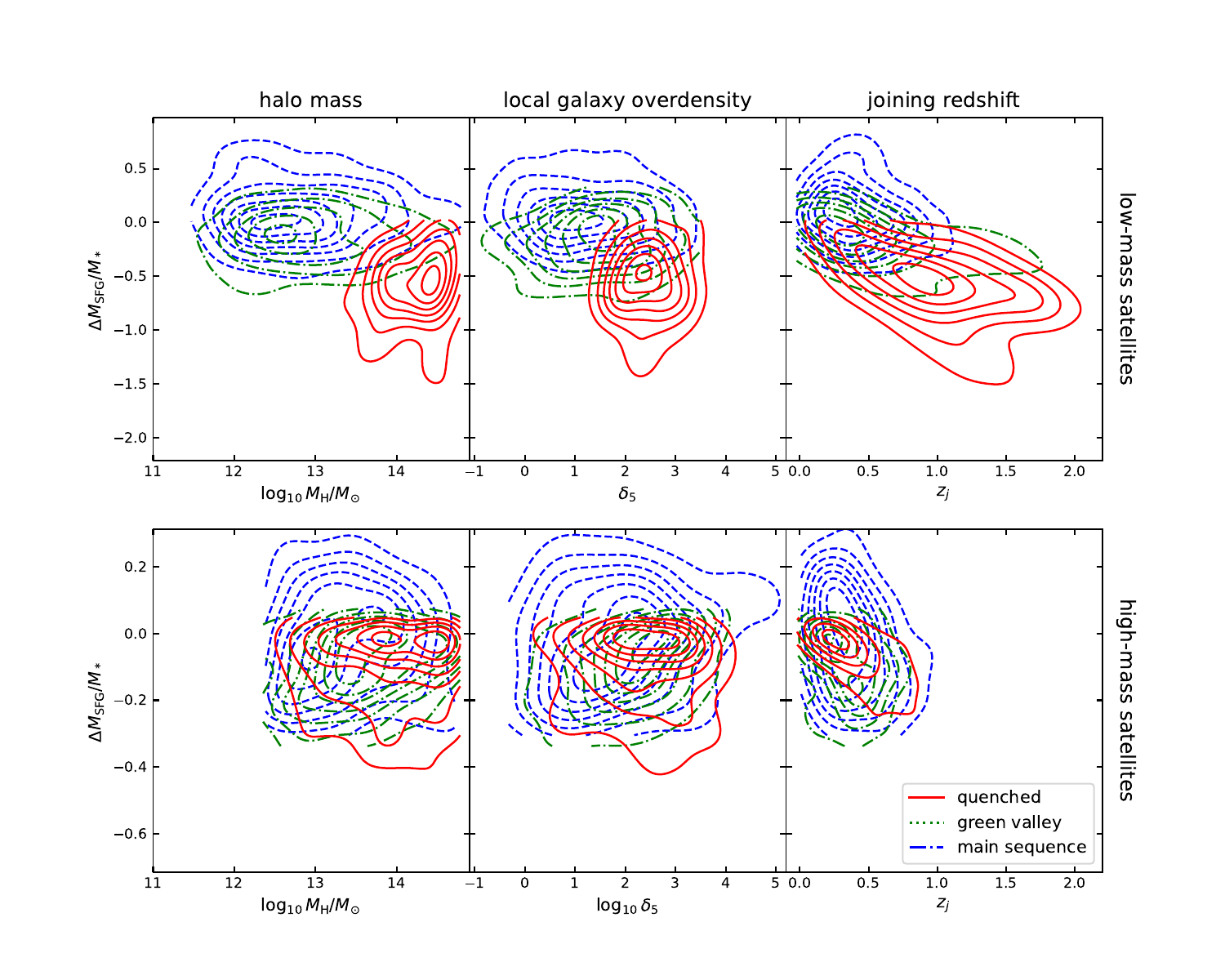}
    \caption{Correlation between \sfgas{} and environmental parameters for low-mass (top) and high-mass (bottom) satellite galaxies: halo mass (left), local galaxy overdensity (middle), and joining redshift (right). Density contours for the main sequence (blue), green valley (green), and quenched (red) populations are shown. Note that the ordinate limits differ between rows.}
    \label{fig:orbitcorr}
  
\end{figure}

In Figure \ref{fig:sat_dist}, we present the correlations between parameters for satellite galaxies. For parameters that are also measured in central galaxies, the plotted ranges are identical to \ref{fig:cent_dist}. For satellite galaxies, we also include the $z_\mathrm{j}$ and \sfgas{} parameters. As for the central galaxies, there is a  correlation between stellar mass and black hole properties (mass and energy injected), as well as a loose $M_*-$\bulge{} correlation. Here, halo mass is no longer correlated with stellar mass, since satellites reside within the larger halo of their central host galaxy. However, halo mass is correlated with $\delta_5$. The redshift at which a satellite joined its host, $z_j$, and the change in star-forming gas, \sfgas, are not well correlated with any of the other parameters and are only slightly correlated with one another. The most negative values of \sfgas{} only occur at the highest halo masses.

Differences between the one-dimensional histograms in Figures \ref{fig:cent_dist} and \ref{fig:sat_dist} are especially illuminating. While only $\approx 6\%$ of quenched TNG100 central galaxies have $M_*/M_\odot <10^{10.4}$, just over $50\%$ of quenched TNG100 satellites fall within this mass regime. The distributions of $M_*$, $M_{BH}$, and \cumQM{} for quenched satellite galaxies appear similar to the distributions of those parameters for quenched central galaxies, with an additional component at low values. This indicates that, while satellite galaxies will undergo mass quenching at high stellar masses, an additional quenching pathway is available for satellite galaxies. 

This additional pathway allows for the quenching of low-mass satellite galaxies for which AGN feedback is weak or non-existent, and is most likely related to environmental processes. Quenched satellite galaxies are more likely to reside in more massive halos and in more dense areas, as shown in the distributions of $M_{H}$ and $\delta_5$ in Figure \ref{fig:sat_dist}. The net effect of environmental processes is on the availability of star-forming gas. 

The bottom right panel in Figure \ref{fig:sat_dist} shows the distribution of \sfgas. When \sfgas{} is positive, star-forming gas has been added to the galaxy, likely through compression shocks or possibly accretion, resulting in main sequence galaxies at $z=0$. In contrast, all quenched galaxies have values of \sfgas that are zero or negative, indicating that these systems entered their hosts' halos as objects that were already quenched, or they were quenched as the halo stripped them of their star-forming gas. %best way to say this last sentence isn't coming to me at this moment. can come in quenched, be stripped, be strangulated/ all of the above

As quenching occurs differently in galaxies of different masses, we further explore the correlation between \sfgas{} and $M_H$ (left), $\delta_5$ (center), and $z_j$ (right) in Figure \ref{fig:orbitcorr} for low- and high- mass satellites separately. From this figure, it is clear that the quenched (red) population of high-mass satellites (bottom row) differs very little from the main sequence (blue) or green valley (green) high-mass satellite populations in these parameter spaces. This is in contrast to the population of quenched low-mass satellites (top row). This indicates that quenching in high-mass satellite galaxies is not driven by environmental factors that remove star-forming gas, but it is the dominanat mechanism for quenching of low-mass satellites. A likely explanation for this phenomenon is that the greater gravitational potential of high-mass satellite galaxies better enables these systems to retain their gas reservoir during environmental interactions. The quenching in high-mass satellites is then primarly driven by AGN feedback energy, which introduces turbulence, reducing star formation efficiency. 

Quenched, low-mass satellite galaxies form the most distinct population in $M_H-$\sfgas{} parameter space in Figure \ref{fig:orbitcorr}. Galaxy-galaxy interactions that scale with local galaxy overdensity could also drive star-forming gas loss, although it is correlated with halo mass (Figure \ref{fig:sat_dist}). Low-mass satellites with high $z_j$ are much more likely to be in the quenched or green valley populations, and $z_j$ is loosely correlated with \sfgas{} loss.

\section{Summary of gradients and global averages}
\label{sec:app_summary}
\begin{figure}[ht!]
    \centering
    \includegraphics[width=\textwidth, keepaspectratio]{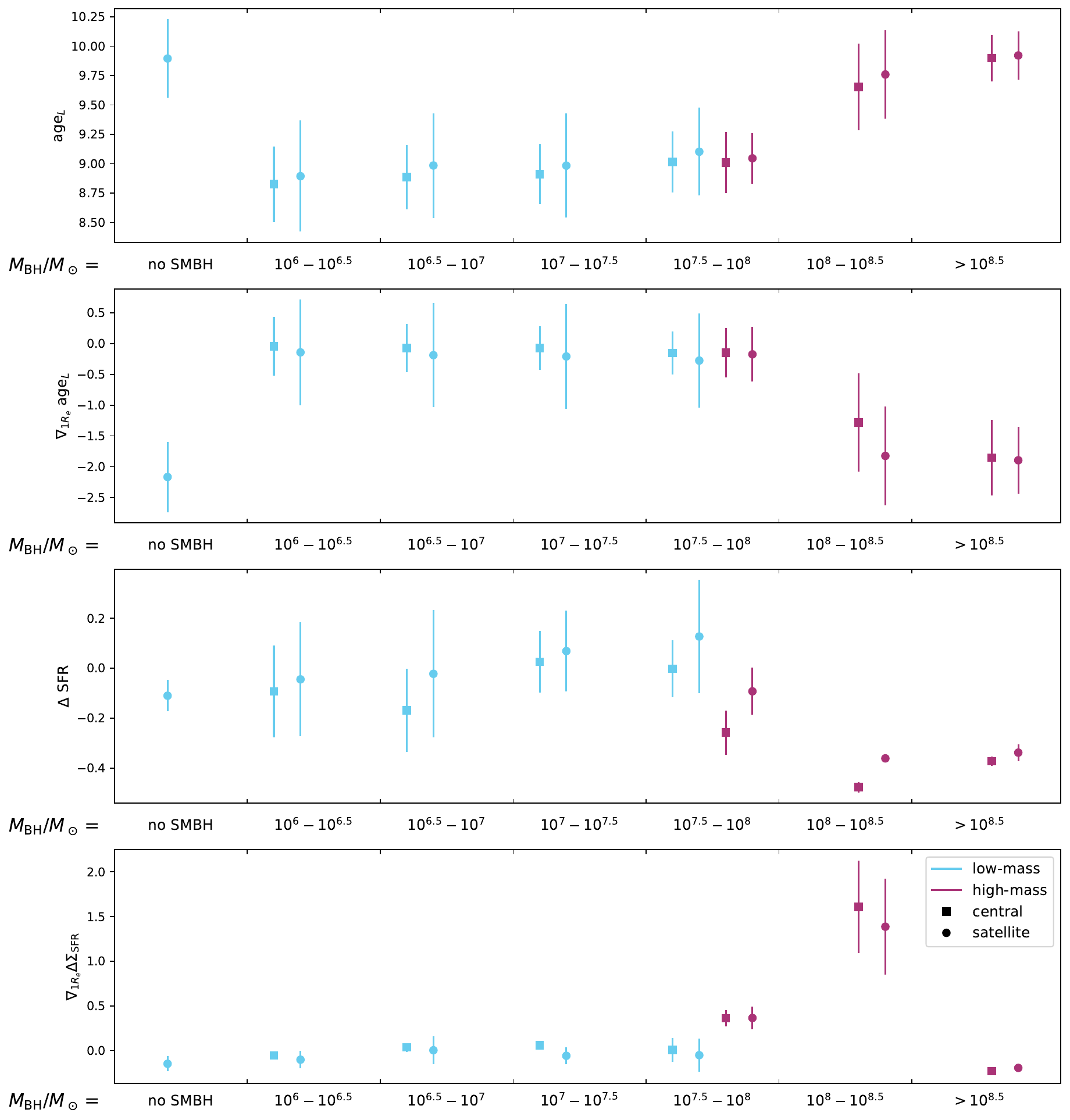}
    \caption{\textnormal{A summary of the median \age{} and SFR, as well as the  gradients measured for populations of TNG100 galaxies whose radial profiles are shown in Figures \ref{fig:BHmass} and \ref{fig:SFR_BHMass}. From top to bottom, the panels are the median global \age{} of each population subset, the \age{} gradient reported in the legend of Figure \ref{fig:BHmass}, the median offset from the global star-forming main sequence ($\Delta$SFR), and the \SFR{} gradients reported in the legend of Figure \ref{fig:SFR_BHMass}. The galaxy populations are divided into bins of $M_{\rm BH}/M_\odot$, which are given by the x-axis. Galaxies in these bins are further divided into high- and low- mass (magenta and blue, respectively) and classification as central or satellite galaxies (square and circle, respectively). Points within each bin are arbitrarily offset for visibility. }
    %For each panel, the variance between low-mass galaxies is slight (with the exception of the case where no SMBH was seeded). However, there is a trend in high-mass galaxies as the mass of the SMBH increases, including older ages, more negative \grad \age, lower $\Delta$SFR, and a steeper \grad \SFR{} for galaxies with $10^6 <M_{\rm BH}/M_\odot < 10^{6.5}$.
    }
    \label{fig:sum_BHmass}
  
\end{figure}

\begin{figure}[ht!]
    \centering
    \includegraphics[width=\textwidth, keepaspectratio]{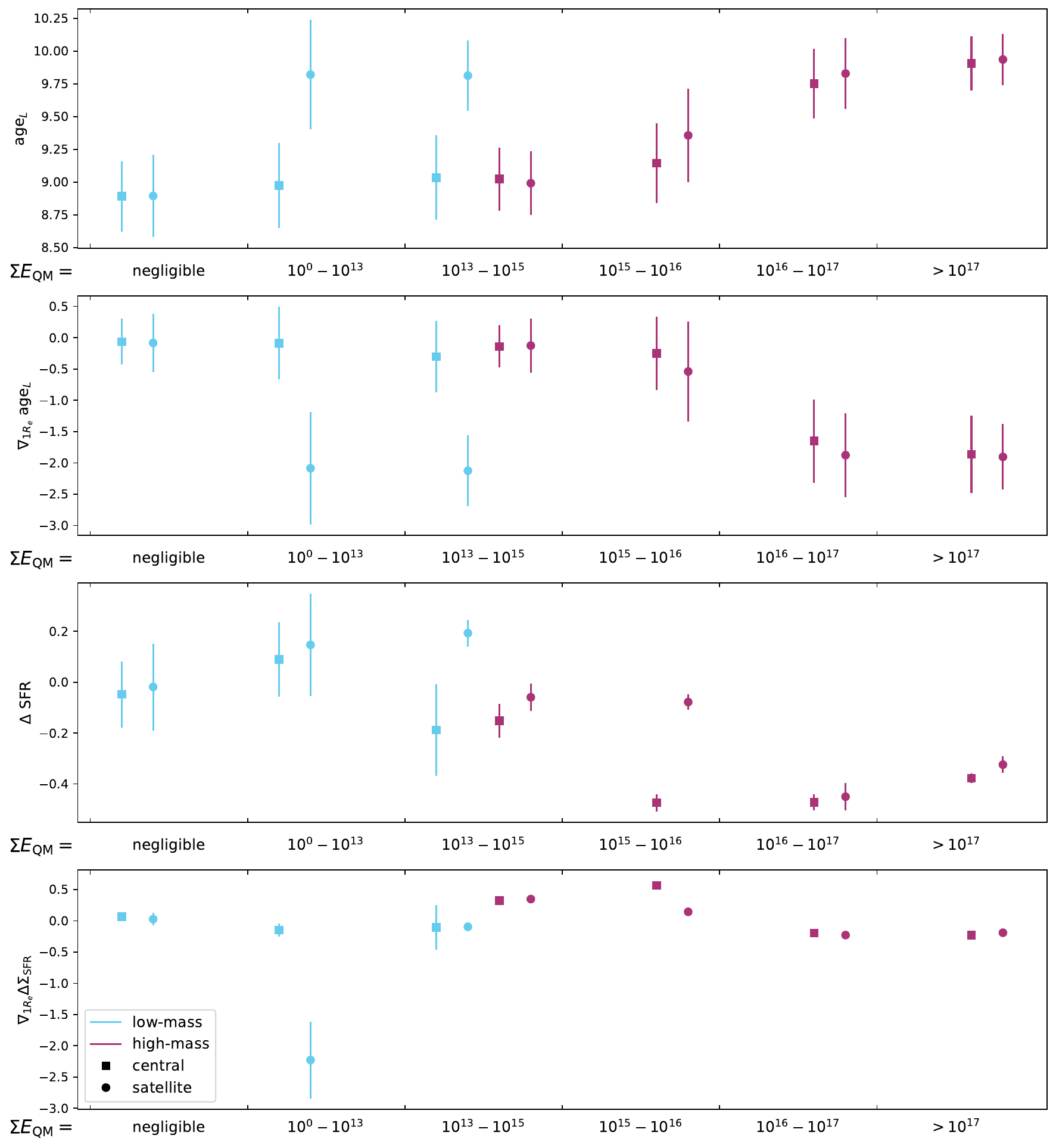}
    \caption{\textnormal{Formatting is identical to Figure \ref{fig:sum_BHmass}, except the figure summarizes results from the \cumQM{} parameter shown in Figures \ref{fig:cumQM} and \ref{fig:SFR_cumQM}.} %For low-mass galaxies, there is little variation among the parameters shown in the four panels, with the exception of low-mass satellites. The cause of this is unclear. For high-mass galaxies, the median \age{} of the populations increase with increasing \cumQM, while the \grad \age{} decreases. The median value of $\Delta$SFR decreases at higher values of \cumQM, until it is below the resolution limit. The steepest gradient occurs in the population with the second-highest bin of \cumQM. This indicates that this population is undergoing active quenching, while the population in the highest \cumQM{} bin are almost entirely quenched.
    }
    \label{fig:sum_cumQM}
  
\end{figure}

\begin{figure}[ht!]
    \centering
    \includegraphics[width=\textwidth, keepaspectratio]{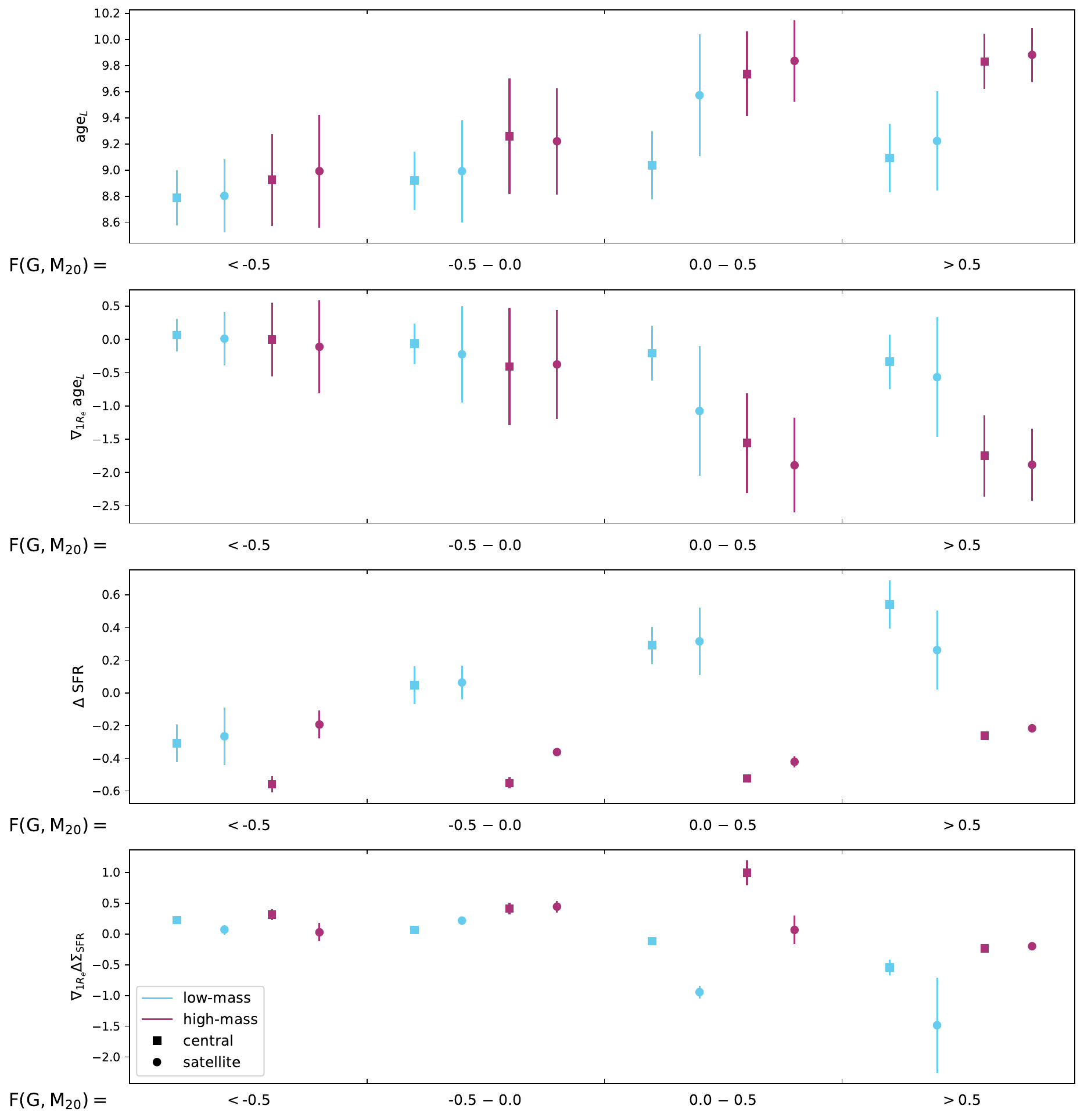}
    \caption{\textnormal{Formatting is identical to Figure \ref{fig:sum_BHmass}, except the figure summarizes results from the \bulge{} parameter shown in Figures \ref{fig:bulge} and \ref{fig:SFR_bulge}.} %There is a large variation in the median \age{} and the measured \grad \age.  
    }
    \label{fig:sum_bulge}
  
\end{figure}

\begin{figure}[ht!]
    \centering
    \includegraphics[width=\textwidth, keepaspectratio]{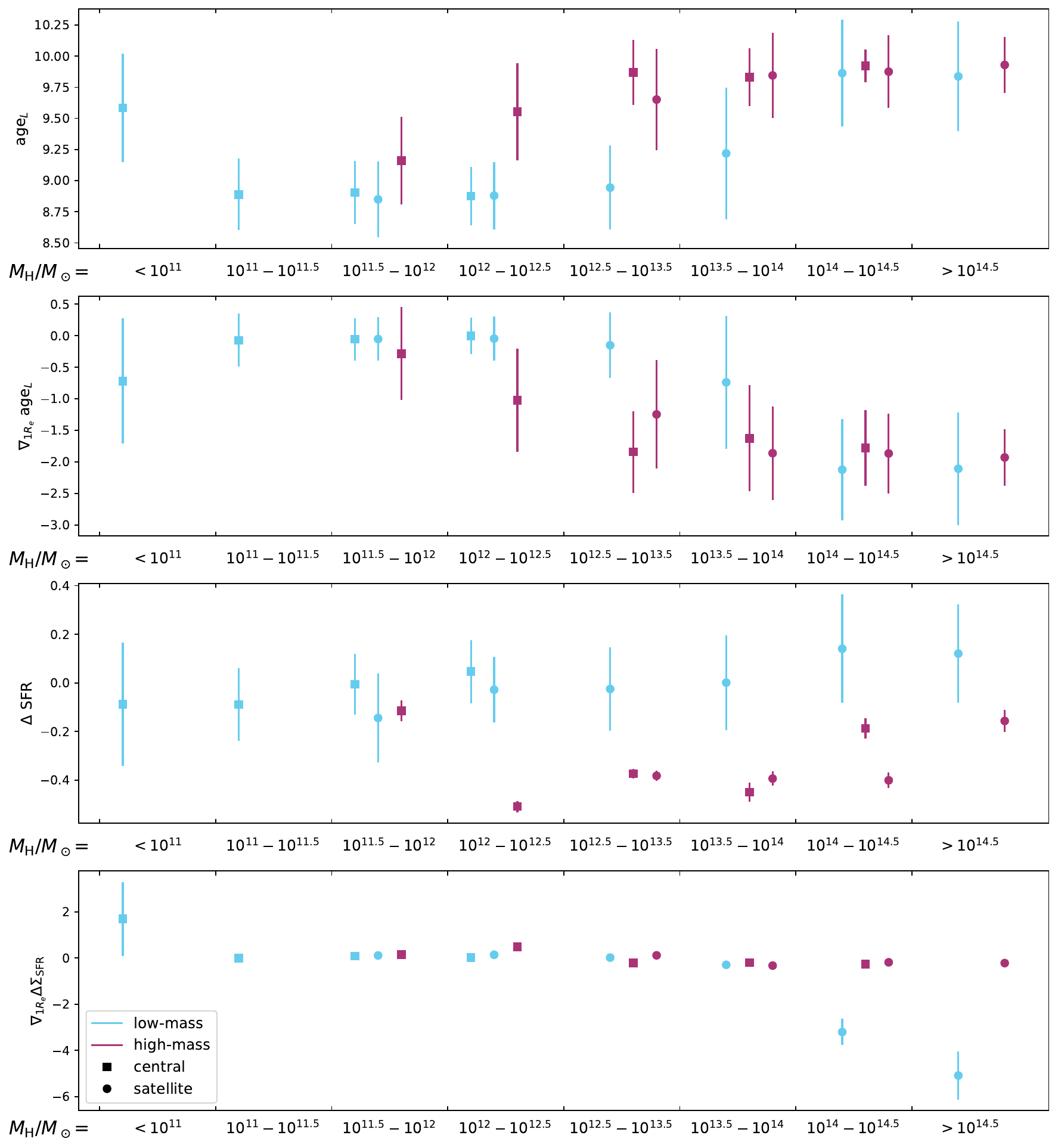}
    \caption{\textnormal{Formatting is identical to Figure \ref{fig:sum_BHmass}, except the figure summarizes results for populations divided by $M_{\rm H}$ shown in Figures \ref{fig:halo} and \ref{fig:SFR_halo}.}}
    \label{fig:sum_Mhalo}
  
\end{figure}

\begin{figure}[ht!]
    \centering
    \includegraphics[width=\textwidth, keepaspectratio]{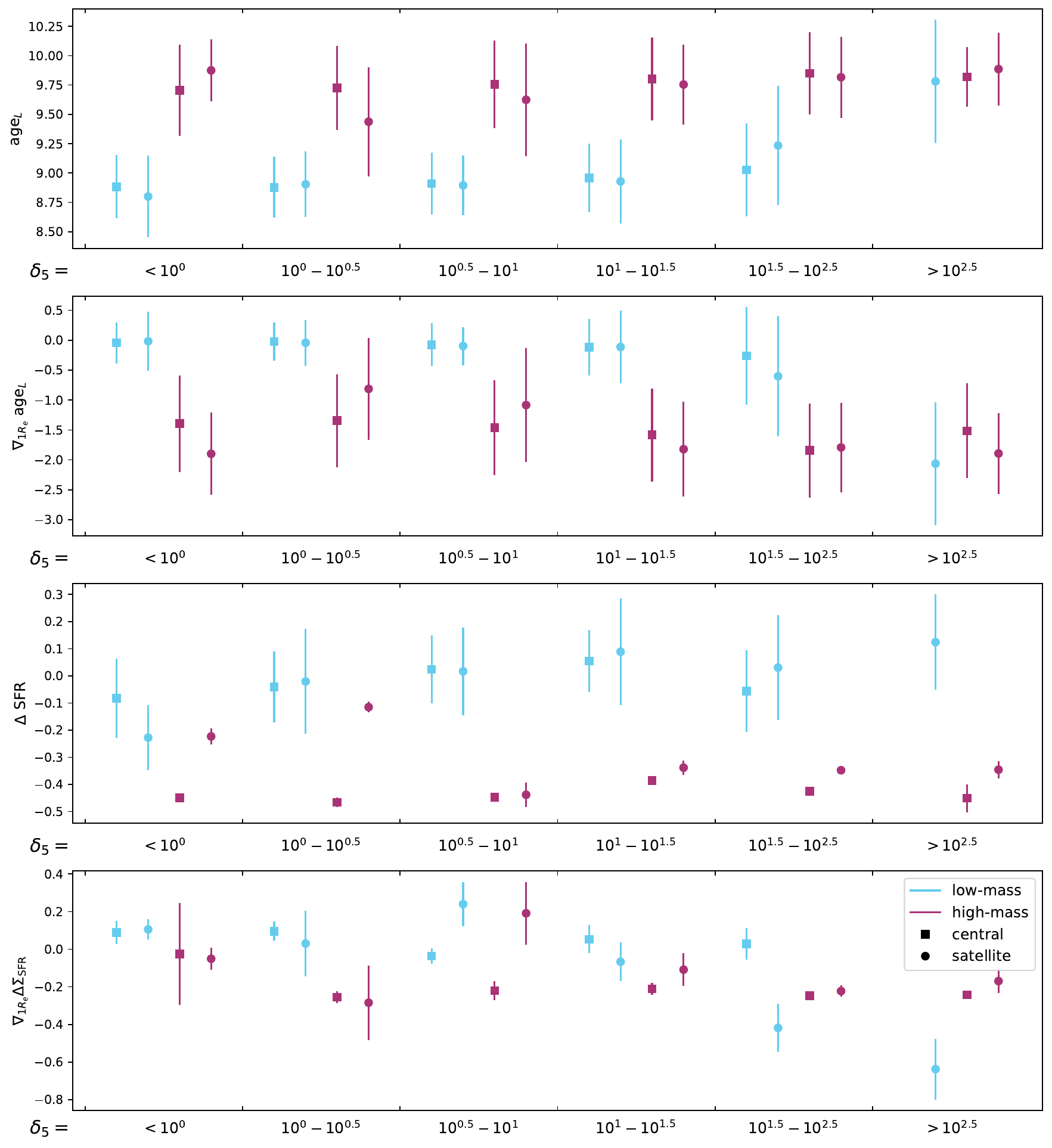}
    \caption{\textnormal{Formatting is identical to Figure \ref{fig:sum_BHmass}, except the figure summarizes results for populations divided by $\delta_5$ shown in Figures \ref{fig:overdens} and \ref{fig:SFR_overdens}.}}
    \label{fig:sum_overdens}
  
\end{figure}

\begin{figure}[ht!]
    \centering
    \includegraphics[width=\textwidth, keepaspectratio]{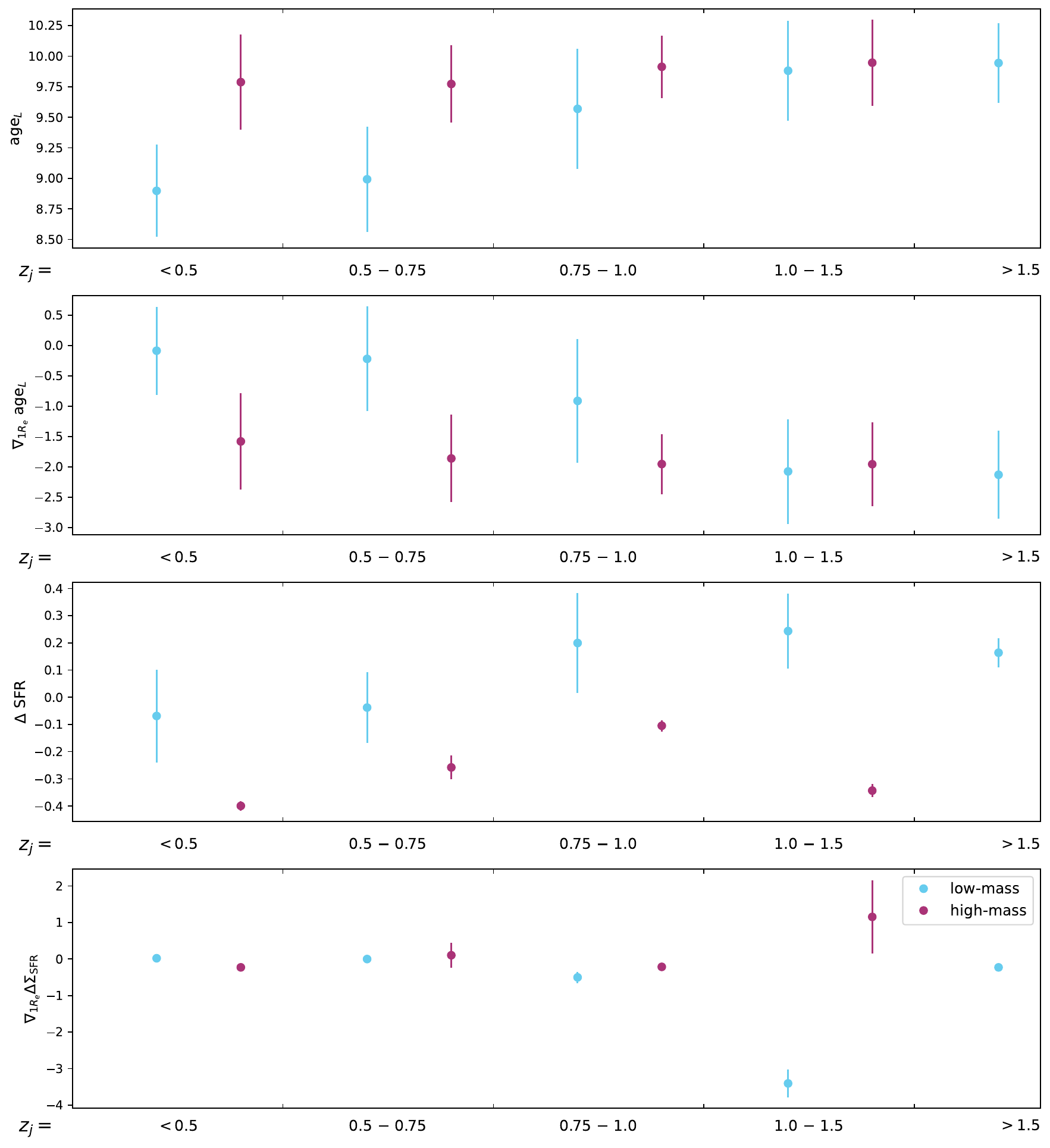}
    \caption{\textnormal{Formatting is identical to Figure \ref{fig:sum_BHmass}, except the figure summarizes results for populations divided by $z_j$ shown in Figures \ref{fig:joinz} and \ref{fig:SFR_joinz}.}}
    \label{fig:sum_joinz}
  
\end{figure}

\begin{figure}[ht!]
    \centering
    \includegraphics[width=\textwidth, keepaspectratio]{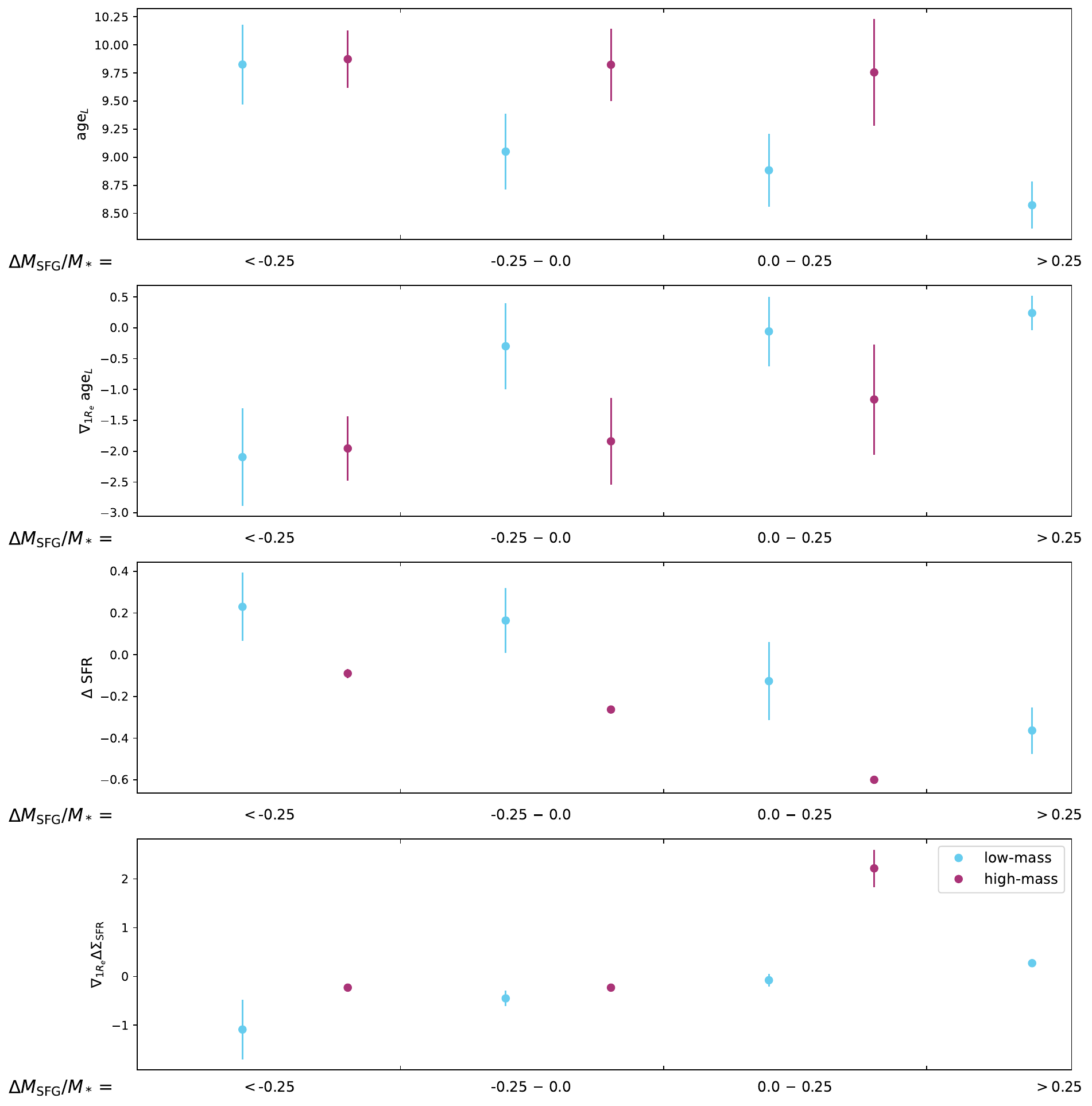}
    \caption{\textnormal{Formatting is identical to Figure \ref{fig:sum_BHmass}, except the figure summarizes results for populations divided by \sfgas{} shown in Figures \ref{fig:SFgas} and \ref{fig:SFR_SFgas}.}} 
    \label{fig:sum_SFgas}
  
\end{figure}

\textnormal{In this appendix, we include figures that summarize the gradients reported in the legends of Figures \ref{fig:BHmass} - \ref{fig:SFR_SFgas} and the overall normalization of the profiles. For the normalization of \age, we take the luminosity-weighted age of all stellar particles that are bound to the galaxy. For the normalization of \SFR, we report $\Delta$SFR, the logarithmic distance from the star-forming main sequence line reported in \cite{McDonough23}. Figures \ref{fig:sum_BHmass} - \ref{fig:sum_SFgas} each summarize the results from one of the explored parameters: $M_{\rm BH}$, \cumQM, \bulge, $M_{\rm H}$, $\delta_5$, $z_j$, and \sfgas. The x-axes are the bins used in Figures \ref{fig:BHmass} - \ref{fig:SFR_SFgas}, and points within those bins are arbitrarily offset for visibility. Low-mass galaxies are given in blue and high-mass galaxies are given in magenta. Central galaxies are shown as squares and satellite galaxies are shown as circles.  From top to bottom, the panels are the median global \age{} for the population with errors given by the standard deviation, the \age{} gradient for the population with errors given by the fit, the median $\Delta$SFR for the population with errors given by the standard deviation, and the gradient of \SFR{} with errors given by the fit. }

\textnormal{These figures do not capture the full variation in the shape of the radial profiles, but provide an overview of our results. The gradients reported are only measured from the center to $1R_e$, and therefore do not capture differences in the outer regions or between the disk and bulge regions.}

\section{Mass-dependency of normalizations}
\label{sec:app_mass}

\begin{figure}[ht!]
    \centering
    \includegraphics[ keepaspectratio]{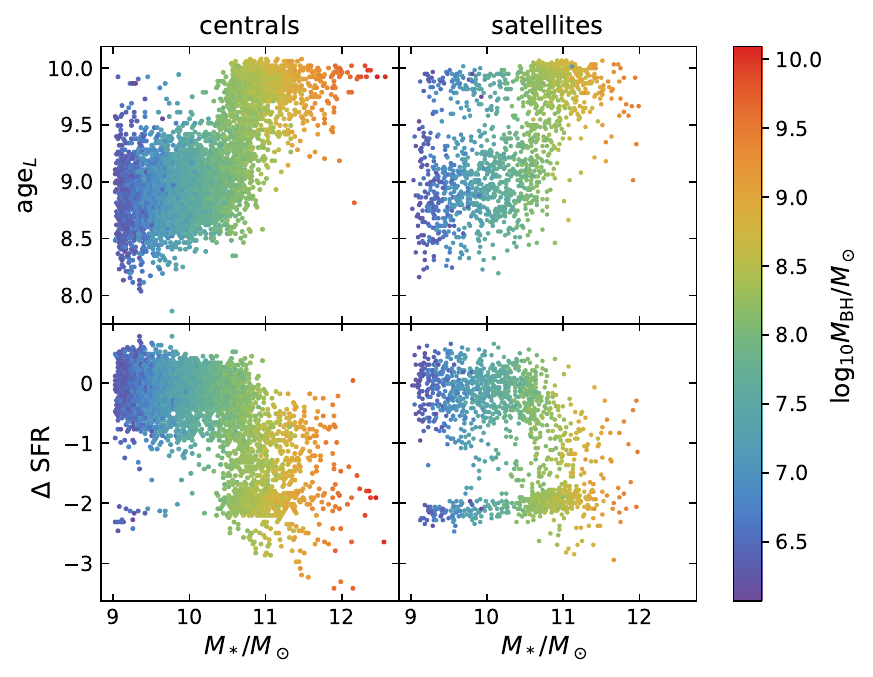}
    \caption{\textnormal{The total \age{} for a galaxy and the galaxy's logarithmic offset from the star-forming main sequence ($\Delta$SFR) is plotted against stellar mass in the top and bottom panels, respectively. Central and satellite galaxies are plotted separately in the left and right columns, respectively. Points are colored according to the SMBH mass, $M_{\rm BH}$.}
    }
    \label{fig:mass_BHMass}
  
\end{figure}

\begin{figure}[ht!]
    \centering
    \includegraphics[keepaspectratio]{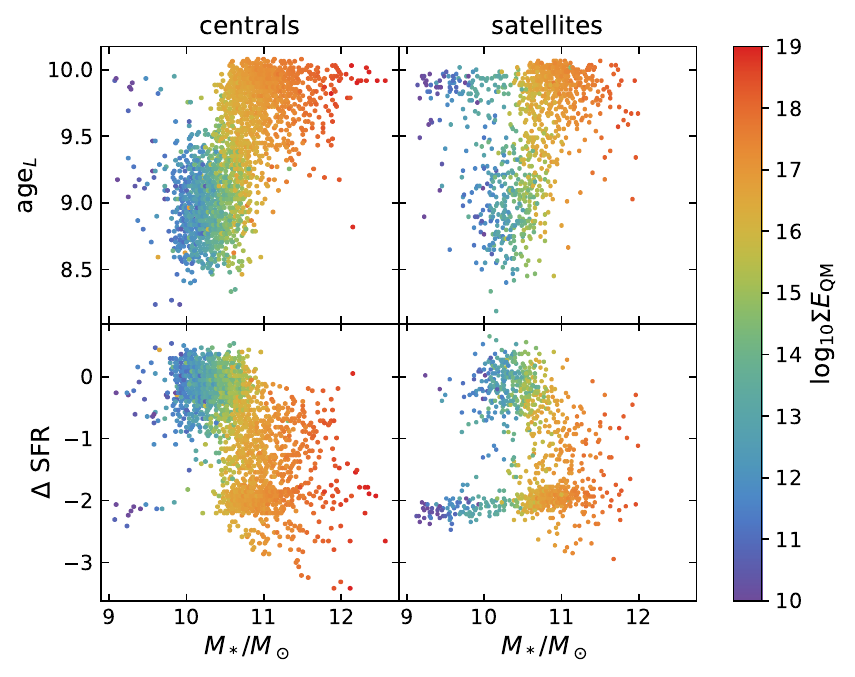}
    \caption{\textnormal{Formatting is identical to Figure \ref{fig:mass_BHMass}, except points are colored by \cumQM.}}
    \label{fig:mass_cumQM}
  
\end{figure}

\begin{figure}[ht!]
    \centering
    \includegraphics[keepaspectratio]{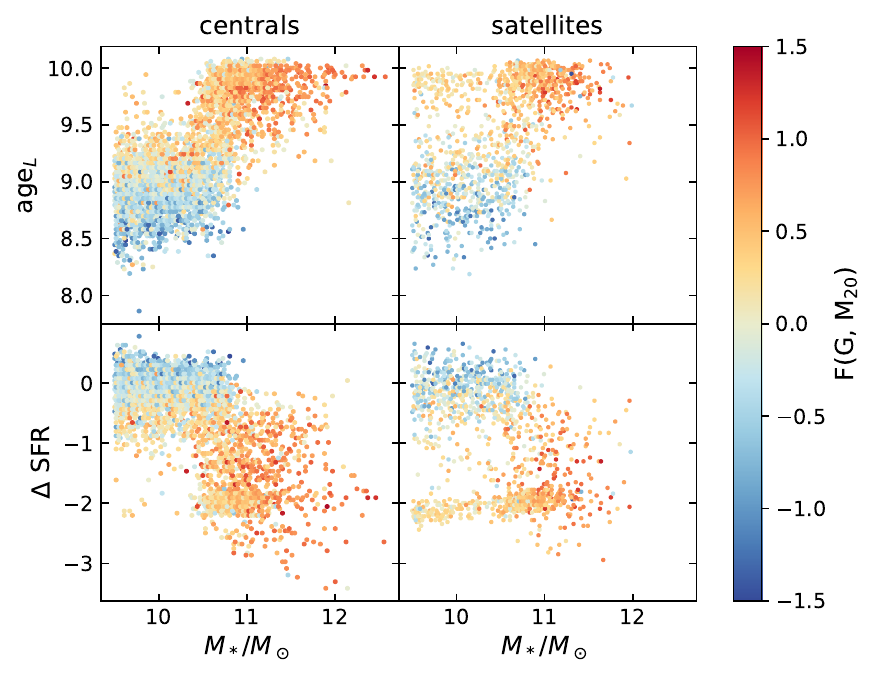}
    \caption{\textnormal{Formatting is identical to Figure \ref{fig:mass_BHMass}, except points are colored by \bulge.}}
    \label{fig:mass_bulge}
  
\end{figure}

\begin{figure}[ht!]
    \centering
    \includegraphics[ keepaspectratio]{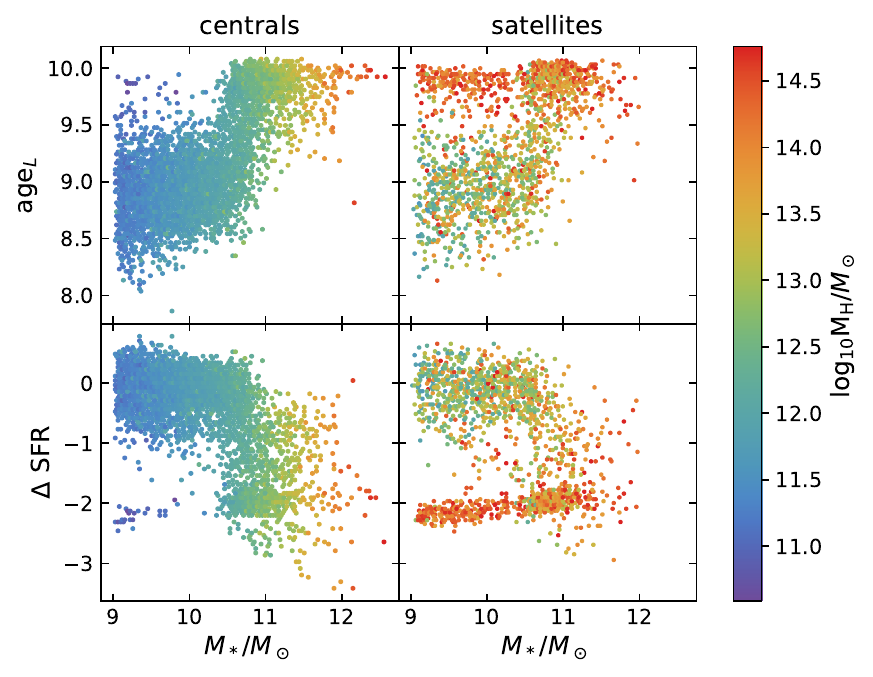}
    \caption{\textnormal{Formatting is identical to Figure \ref{fig:mass_BHMass}, except points are colored by $M_{\rm H}$.}}
    \label{fig:mass_Mhalo}
  
\end{figure}

\begin{figure}[ht!]
    \centering
    \includegraphics[keepaspectratio]{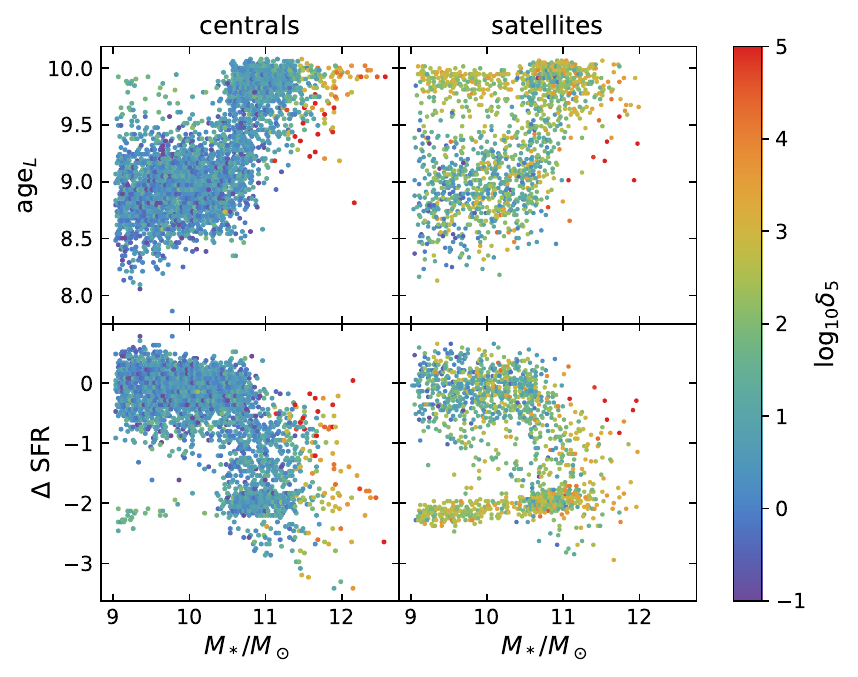}
    \caption{\textnormal{Formatting is identical to Figure \ref{fig:mass_BHMass}, except points are colored by $\delta_5$.}}
    \label{fig:mass_overdens}
  
\end{figure}

\begin{figure}[ht!]
    \centering
    \includegraphics[keepaspectratio]{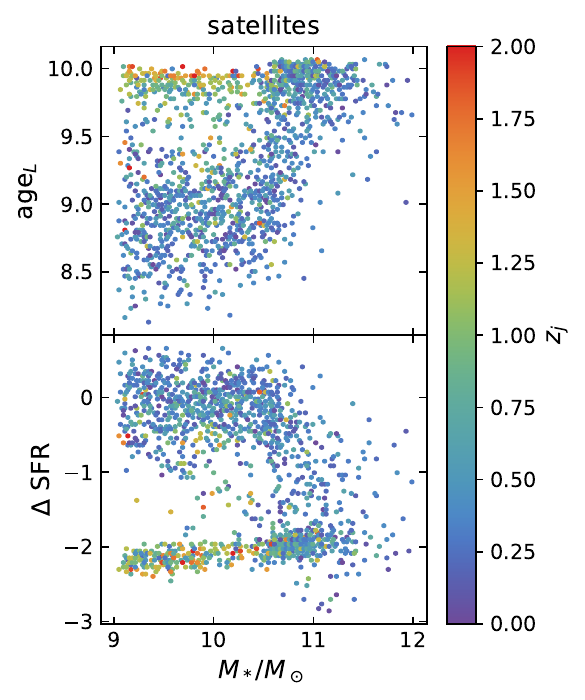}
    \caption{\textnormal{Formatting is identical to the right column of Figure \ref{fig:mass_BHMass}, except points are colored by $z_{\rm j}$.}}
    \label{fig:mass_joinz}
  
\end{figure}

\begin{figure}[ht!]
    \centering
    \includegraphics[ keepaspectratio]{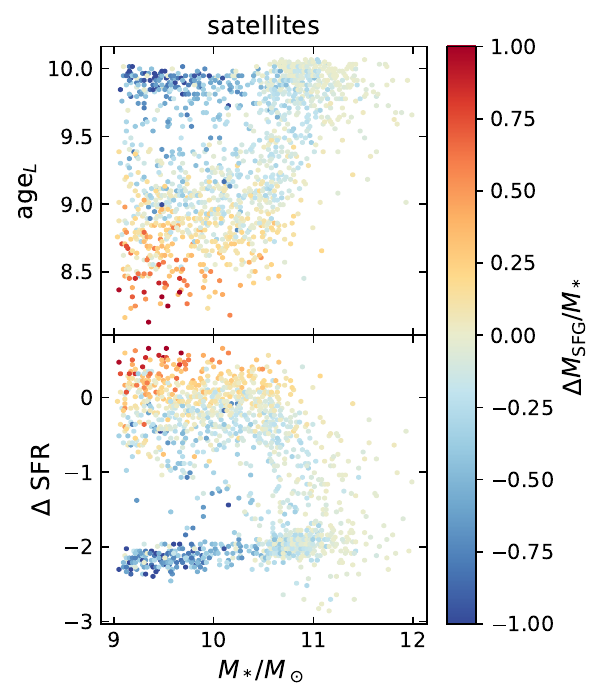}
    \caption{\textnormal{Formatting is identical to Figure \ref{fig:mass_joinz}, except points are colored by \sfgas.}}
    \label{fig:mass_SFgas}
  
\end{figure}

\textnormal{In this article, we have divided our sample into high- and low- mass galaxies (\himass{} and \lomass, respectively). However, the shape and normalization of \age{} and \SFR{} profiles are mass-dependent (e.g., Figures 4 and 5 in \citealt{McDonough23}). Many of the parameters we have explored in \S \ref{sec:inprof} and \S \ref{sec:envprof} are intrinsically related to stellar mass, which may modulate the effect. In this appendix, we present figures that explore how the cumulative \age{} and the logarithmic offset from the global star-forming main sequence ($\Delta$SFR) for galaxies in our sample depend on stellar mass and the explored parameters. In Figures \ref{fig:mass_BHMass} - \ref{fig:mass_overdens}, we plot $M_*$ against total \age{} (top rows) and $\Delta$SFR (bottom rows) for central galaxies (left panels) and satellites (right panels). Points are colored by the value of the explored parameters for each galaxy. Figures \ref{fig:mass_joinz} and \ref{fig:mass_SFgas} are identical in formatting, but only include a single column for satellite galaxies. While we use a similar color scheme, the point colors do not necessarily correspond to the colors of profiles in Figures \ref{fig:BHmass} - \ref{fig:SFR_SFgas}. In Figures \ref{fig:mass_bulge} and \ref{fig:mass_SFgas}, we use a different color scheme that diverges at \bulge$=0$ and \sfgas$=0$ to emphasize the difference between galaxies with and without bulges and satellites that have lost and gained star-forming gas. We remind readers that low-levels of SFR in the simulation are not resolved, but are here artificially set to $sSFR = 10^{-12} {\rm yr}^{-1}$ with some scatter, which results in the linear trend appearing in some panels at $\Delta {\rm SFR} \sim -2$. We do not include figures that explore the mass-dependency of profile gradients because the profiles of individual galaxies are too noisy to obtain good estimates of the slopes, especially at lower masses where there are less stellar particles.}

%% This command is needed to show the entire author+affiliation list when
%% the collaboration and author truncation commands are used.  It has to
%% go at the end of the manuscript.
%\allauthors

%% Include this line if you are using the \added, \replaced, \deleted
%% commands to see a summary list of all changes at the end of the article.
%\listofchanges

\end{document}